\documentclass{ws-ijmpd}
\usepackage[super]{cite}
\usepackage{xcolor}
\usepackage[verbose,hypertexnames=false]{hyperref}
\hypersetup{colorlinks=false,allbordercolors=blue,pdfborderstyle={/S/U/W 1}}
\usepackage{physics}
\begin{document}

\markboth{Kapil Goswami and Peter Schmelcher}{A Unified Local Light-shifts Encoding For Solving Optimization Problems on a Rydberg Annealer}

%
\catchline{}{}{}{}{}
%

\title{A Unified Local Light-shifts Encoding For Solving Optimization Problems on a Rydberg Annealer}

\author{Kapil Goswami}

\address{Zentrum f\"ur Optische Quantentechnologien, Universit\"at Hamburg, Luruper Chaussee 149, 22761 Hamburg, Germany\\
kapil.goswami@uni-hamburg.de}

\author{Peter Schmelcher}

\address{Zentrum f\"ur Optische Quantentechnologien, Universit\"at Hamburg, Luruper Chaussee 149, 22761 Hamburg, Germany\\ 
peter.schmelcher@uni-hamburg.de}

\maketitle


\begin{abstract}
Combinatorial optimization problems play a central role in computer science with many real world applications. A number of relevant problems remain computationally difficult to solve as they lie in the NP-hard complexity class. We present a unified framework for solving such optimization problems represented in the quadratic unconstrained binary optimization (QUBO) formalism, namely two-SAT, XOR-SAT, mixed-two-XOR-SAT, set packing, quadratic assignment, binary clustering, and protein folding, by expanding the domain of applications of \textit{PRR, 6(2), 023031}. A direct mapping from the QUBO form of these problems onto the Rydberg quantum platform is demonstrated as our first step. This mapping to the Rydberg system depends on distance-dependent long-range interactions and configurable local detuning, thus reducing resource overhead and improving scalability. 
Following-up on the encoding, the solution is reached by steering the system toward the ground state of the target Hamiltonian using an optimized quantum annealing protocol that controls the time-dependent detuning and Rabi frequency profiles. The framework can handle a variety of problems, each with different complexity. To quantify the complexity of any problem, a generalized hardness parameter is introduced that compares different problems based on the structure of their optimization landscapes.
This is a proceedings contribution to the Athens Workshop in Theoretical Physics: 10th Anniversary, held at the National and Kapodistrian University of Athens on December 17-19 2025.
\end{abstract}


\section{Introduction}	

Developing quantum algorithms to solve combinatorial optimization problems is a growing field which is extensively explored \cite{cho1998fast,hochbaum1998analysis,perelshtein2022practical,butenko2003maximum,scheideler2008log,goswami2024integer,goswami2026solving}. 
This is due to the fact that optimization problems are generally hard to solve using classical algorithms (NP-hard complexity class) while having many practical applications \cite{karp1975computational,papadimitriou1988optimization,scheideler2008log,perelshtein2022practical}. Hence, these problems serve the purpose for testing the computational advantage of quantum algorithms over classical ones and provide insights to help benchmark both the computational platforms, viz., quantum and classical \cite{mandal2023review,abbas2024challenges,pirnay2024principle,goswami2025qudit}. Extending the approach developed in Ref.~\refcite{kapil}, we encode and solve the following problems, namely, two-SAT, XOR-SAT, mixed-two-XOR-SAT, set packing, quadratic assignment problem, binary clustering, and protein folding \cite{wolsey1999integer,gu1997algorithms,hochba1997approximation,cela2013quadratic,zhang2018binary,compiani2013computational}.  

One crucial point is that all the mentioned problems can be represented in quadratic unconstrained binary optimization (QUBO) form, which is native for encoding them using Ising spins \cite{barahona1988application,glover_tutorial_2019,lucas_ising_2014}. These problems are chosen because of their various applications and varying computational complexity. Out of all the examples, the two-SAT and two-XOR-SAT problems can be solved in polynomial time using classical algorithms, while the rest of the problems are NP-hard \cite{schaefer1978complexity,fraenkel1993complexity,karp1972complexity}. Since the problems require a mapping to the quantum system, their structure plays an important role in determining the resources needed for encoding them. For instance, the QUBO representation of the protein folding and quadratic unconstrained assignment problems scales quadratically with the number of variables, making them expensive to encode as compared to the other problems with a similar size \cite{irback2022folding,ayodele2022penalty,nusslein2022algorithmic}. The number of variables in the QUBO directly translates to the number of spins in the Ising Hamiltonian, which corresponds to the number of physical qubits required in the quantum platform \cite{lucas_ising_2014}. Apart from the scaling of QUBO with the problem size, another layer of encoding, that is, Ising spins to the physical qubits, poses a challenge as it needs to be efficient as well \cite{mbeng2024quantum,lee2026fundamental}. 

Currently, the properties of physical qubits are heavily dependent on the corresponding quantum computing platform. Superconducting qubits and trapped ions platforms can be used for solving these problems; however, there are issues regarding their scalability and sensitivity \cite{devoret2004superconducting,harter2014long,bruzewicz2019trapped}. Recent neutral Rydberg atom experiments with hundreds of physical qubits, precise control over interactions between them, and the potential for scalability \cite{browaeys2020many,Gal,gross2017quantum,evered2023high,bluvstein2024logical}, motivate us to utilize them in this work. Furthermore, research on methods for finding optimal solutions and encoding schemes for the problem has provided us with many choices for solving a QUBO using Rydberg atoms \cite{ebadi2022quantum,kim_rydberg_2021,kapil}. Among the various methods, variational algorithms are particularly popular; however, they often suffer due to the large number of parameters involved, including the initial state ansatz, the choice of mixer, and the number of layers \cite{farhi_quantum_2014, hadfield_quantum_2019,peruzzo2014variational}. Additionally, optimizing these parameters is known to be NP-hard \cite{bittel2021training, bittel2022optimizing}. The complexity of the variational algorithms further increases with the number of qubits (or atoms) required to encode the problem, which is often the case with Rydberg blockade-based encoding schemes \cite{urban2009observation,ebadi2022quantum,kim_rydberg_2021}. The blockade-based schemes generally have a quadratic overhead in the number of atoms when encoding QUBO to a Rydberg platform \cite{nguyen_quantum_2022}. For bypassing these issues, we use a non-blockade-based encoding and propose an optimized quantum annealing protocol to solve these optimization problems on the Rydberg platform. This approach was originally developed in Ref.~\refcite{kapil} for max-cut and maximum independent set problems, and we extend it to other optimization problems. The listed problems have not been treated in a unified manner for encoding, whereas our scheme can be applied to any QUBO using the Rydberg atoms.

As mentioned, in this work, we encode and solve QUBO problems on a Rydberg platform by employing a direct mapping to spin models using localized light shifts on individual atoms. 
The quantum annealing is then optimized by defining a suitable, physically viable objective function and using a combination of gradient-free along with gradient based optimizers to navigate any optimization landscape \cite{kadowaki1998quantum,polak2012optimization,hasdorff1976gradient,conn2009introduction,hare2013survey}. For the Rydberg platform, the detuning and the Rabi frequency profiles are tuned as a function of time to steer the system to the ground state of the target Hamiltonian representing the QUBO problem. We apply this scheme to a range of prototypical problems with $3$ to $5$ variables, achieving solutions with an approximation ratio of approximately $0.98$ to $0.999$. To benchmark each problem, we define a generalized hardness parameter that considers various factors within the problem structure contributing to its difficulty. We also conduct a comparative study for all these problems using the hardness parameter as a measure to analyze them.

The outline of the manuscript is as follows. In Section~\ref{Rydberg}, we briefly provide the theoretical description of the Rydberg simulator and the corresponding Ising representation of the Rydberg Hamiltonian. The definitions of the optimization problems with their spin representation are provided in Section~\ref{Problem}. Then, we outline the encoding scheme and the optimal quantum annealing protocol in Section~\ref{LOQAL}. The hardness parameter is discussed and defined in Section~\ref{HP}. The results for the optimal protocols for all the problems using the scheme and the collective analysis of the hardness of the problems are provided in Section~\ref{Results}. Finally, Section~\ref{Conclusions} contains our conclusions and outlook. 

\section{Rydberg Simulator}
\label{Rydberg}
In this work, ultracold atoms trapped in optical tweezers are considered to map the optimization problems \cite{kaufman2021quantum,zeiher2016many,maurer2011spatial}. Each atom encodes the binary information in two energy levels, viz., ground state ($\ket{g}$), and a Rydberg excited state ($\ket{e}$) that are coupled using a laser \cite{gaetan2009observation,low2012experimental}. The setup consists of site-dependent detuning $\Delta_j$ (laser parameter describing the difference between the frequency of the applied field and the natural frequency associated with the atomic transition) and global Rabi frequency $\Omega$ that couples the two states $\ket{g},\ket{e}$ (it is proportional to the intensity of the driving field and the dipole moment associated with the atomic transition). The Hamiltonian expressed in the atomic basis is given as follows
\begin{align}
\hat{H}_{Ryd} = \frac{\Omega}{2} \sum_j \ket{e}_j \bra{g} +\ket{g}_j \bra{e} - \sum_j \Delta _j \ket{e}_j \bra{e}
+ \sum_{k<j} V_{kj}\qty(\ket{e}_k \bra{e} \otimes \ket{e}_j \bra{e})
\label{Ryd0}    
\end{align}
where $V_{kj} = C_6/|\mathbf{r}_j - \mathbf{r}_k|^6$ is the van der Waals ($C_6 > 0$) interaction between the Rydberg atom's excited states $\ket{e}_k,\ket{e}_j$ and $C_6$ is the associated dispersion coefficient. The atoms are labeled as $j,k$ with $\mathbf{r}_j,\mathbf{r}_k$ as the corresponding positions. The above Hamiltonian is represented using the Pauli spin operators in the limit $\Omega = 0$ in the form of the Ising Hamiltonian with a longitudinal field,
\begin{align}
\hat{H}_{Ising} = \sum_{j=1}^N  \qty(-\frac{\Delta_j}{2} + \frac{1}{4} \sum_{k=1, j \neq k}^N V_{jk}) \hat{\sigma}^{z}_{j} 
+ \frac{1}{4} \sum_{j=1}^{N-1} \sum_{k=j+1}^N V_{jk} \hat{\sigma}^{z}_{j} \hat{\sigma}^{z}_{k} + C,
\label{eq:IsingHam}
\end{align}
where the constant $C = \frac{1}{4} \sum_{j=1}^{N-1} \sum_{k=j+1}^{N} V_{jk}$ is just an energy shift. The local longitudinal field depends on the detuning as well as the interaction term $ \sum_{k=1, j \neq k}^N V_{jk}$. The Ising Hamiltonian formulation is useful for encoding optimization problems in the QUBO framework. A few of such problems are provided in the next section, which are solved using the Rydberg simulator in this work.

\section{Problem Definitions}
\label{Problem}
Below, we define the classical cost functions for various exemplar QUBO problems along with their Ising representation. These problems are encoded in the Rydberg Hamiltonian using local detuning via the Ising Hamiltonian formulation.

\subsection{Two-SAT and XOR SAT}
The Boolean satisfiability problem (SAT) is defined over binary variables in a logical formula taking $0/1$ values to satisfy the entire formula \cite{gu1997algorithms}. In general, there can be multiple formulas, and the problem is called \textit{satisfiable} if there exists at least one assignment of variables satisfying all the formulas; otherwise, it is termed \textit{unsatisfiable}. SAT is an NP-complete problem, i.e., there are no known efficient classical algorithms solving the problem in its general form \cite{karp1972complexity}. The problem is crucial for solving complex, real-world problems in hardware/software verification.

\textbf{Two-SAT} problem is a restricted version of the SAT problem in which each clause contains at most two variables (or their negations). Typically, a two-SAT problem consists of clauses such as $(x_i \lor x_j)$ or $(\neg x_k \lor x_l)$, and the goal is to find an assignment of binary variables that satisfies all clauses. The corresponding decision problem is to determine whether the problem is satisfiable or not. Unlike the general SAT, two-SAT is solvable in polynomial time using graph-based methods \cite{schaefer1978complexity}. Despite that, two-SAT is crucial as it captures many real-world constraint systems such as scheduling with pairwise constraints and circuit verification \cite{ganai2007sat}. To represent the problem as QUBO, each clause is converted into a penalty that is zero when satisfied. The cost function is then given as $C_{2SAT} = \sum_{\text{clauses}} P \cdot (\text{unsatisfied condition})$ where $P$ is the penalty of the unsatisfiable clause and the goal is to minimize $C_{2SAT}$.

For example, take two clauses, $(x_i \lor x_j) \rightarrow P(1 - x_i)(1 - x_j)$
and $(x_j \lor x_k) \rightarrow P (1-x_j)(1 - x_k)$, the QUBO representation is $C_{2SAT}(x) = 2P + P(-x_i-2x_j -x_k) + P(x_ix_j + x_jx_k)$.
The corresponding Ising Hamiltonian by using  $x_i = \frac{1 - s_i}{2}$ such that $s_i \in \{-1,+1\}$ and replacing $s_i$ with the Pauli-z operator $\hat{\sigma}^z_i$ is given as,

\begin{equation}
    \hat{H}_{2SAT} = \frac{P}{4} \hat{\sigma}^z_i + \frac{P}{2} \hat{\sigma}^z_j + \frac{P}{4} \hat{\sigma}^z_k + \frac{P}{4} \hat{\sigma}^z_i \hat{\sigma}^z_j + \frac{P}{4} \hat{\sigma}^z_j \hat{\sigma}^z_k + \text{const}.
    \label{2satexH}
\end{equation}
Similarly, any clause can be mapped to an Ising Hamiltonian with a penalty term, where the ground state stores the configuration satisfying the maximum number of clauses. 

\textbf{XOR-SAT} is also a restricted SAT problem where clauses are expressed as exclusive OR (XOR) constraints, such as $x_i \oplus x_j = 1$, also termed as parity constraints. Due to the structure of such constraints, the problem plays a key role in coding theory, cryptography, and error-correcting codes \cite{jordan2025optimization,dietzfelbinger2010tight}. XOR-SAT constraints represent frustrated spin systems and glassy energy landscapes, making it relevant for studying quantum annealing and benchmarking optimization hardware \cite{king2019quantum}.
In the QUBO form of the problem, the penalty term is defined as the square of the constraint. For example, the condition $x_i \oplus x_j = 1$ corresponds to the cost function $C_{XSAT}(x) = (x_i + x_j - 1)^2 \equiv 1 - x_i - x_j + 2x_i x_j$ to be minimized.
The corresponding Ising Hamiltonian is thus, 
\begin{equation}
    \hat{H}_{XSAT} =  \frac{1}{2} \hat{\sigma}^z_i \hat{\sigma}^z_j + \text{const}.
    \label{xsatexH}
\end{equation}

Both problems, two-SAT and two-XOR SAT, are solvable in polynomial time using classical algorithms; however, satisfying a combination of both constraints is NP-hard.  
The \textbf{mixed two-SAT-XOR-SAT problem} combines classical logical satisfiability constraints (two-SAT) with parity-based conditions (XOR-SAT) \cite{im2025accelerating}. The two-SAT enforces clauses of the form $(x_i \lor x_j)$, while the XOR part introduces algebraic constraints that are either equality or inequality between variables. This hybrid structure is particularly important in real-world logical systems as it can simulate higher-order SAT problems \cite{ansotegui2025sat}.
Using the penalty and Hamiltonian construction for the individual problems, the Ising representation of the mixed problem (using Eqs. \ref{2satexH} and \ref{xsatexH}) is given as, 
\begin{equation}
\hat{H}_{m2XSAT} = \frac{P}{4} \hat{\sigma}^z_i + \frac{P}{2} \hat{\sigma}^z_j + \frac{P}{4} \hat{\sigma}^z_k + \frac{P}{4} \hat{\sigma}^z_i \hat{\sigma}^z_j + \frac{P}{4} \hat{\sigma}^z_j \hat{\sigma}^z_k + \frac{1}{2} \hat{\sigma}^z_i \hat{\sigma}^z_j + \text{const}.
\label{msatexH}
\end{equation}

\subsection{Set Packing}
The \textbf{set packing} problem is defined over a set \(U = \{e_1, e_2, \ldots, e_m\}\) from which a collection of \(n\) subsets \(S_1, S_2, \ldots, S_n \subseteq U\), each with weights \(w_i > 0\), are selected such that no two chosen subsets overlap i.e., \(S_i \cap S_j = \emptyset\), while maximizing the number (or total weight) of selected sets \cite{hochbaum1998analysis}. This can be represented as a graph problem, where nodes correspond to subsets and edges encode pairwise overlaps (i.e., an edge exists between the nodes \(i\) and \(j\) if and only if \(S_i \cap S_j \neq \emptyset\)). Set packing is widely used in resource allocation, scheduling, and combinatorial design problems where such overlap constraints arise \cite{vemuganti1998applications}. It is NP-hard; however, the problem maps directly to QUBO (and the Ising Hamiltonian) where interactions encode pairwise conflicts \cite{hazan2006complexity}. To represent the problem with binary variables, each subset is associated with \(x_i \in \{0,1\}\), indicating whether it is selected or not. The goal is to maximize the total weight of the subsets while penalizing overlaps.
The QUBO formulation and the cost function to minimize ($C_{SP}(x)$) is,
\begin{equation}
    C_{SP}(x) = -\sum_i w_i x_i + P \sum_{(i,j)\in D} x_i x_j,
    \label{setpackC}
\end{equation}
where \(w_i\) is the weight of subset \(S_i\), \(P\) is the penalty coefficient for overlapping subsets, and \(D = \{(i,j) : S_i \cap S_j \neq \emptyset\}\) is the set of conflicting pairs. The second term ensures that no two sets with finite overlap of the elements are selected. The corresponding Ising Hamiltonian is thus, 
\begin{equation}
    \hat{H}_{SP} = \sum_{i=1}^n \left( \frac{w_i}{2} - \frac{d_i P}{4} \right) \hat{\sigma}^z_i + \sum_{(i,j) \in D} \frac{P}{4} \hat{\sigma}^z_i \hat{\sigma}^z_j + \text{const},
    \label{setpackH}
\end{equation}
where \(d_i = |\{j : (i,j) \in D\}| = |\{j : S_i \cap S_j \neq \emptyset\}|\) is the degree of node \(i\) in the graph representation, indicating the number of subsets that overlap with \(S_i\).

\subsection{Quadratic Assignment Problem (QAP)}

The \textbf{quadratic assignment problem} (QAP) models the maximum assignment of facilities to locations while minimizing the costs \cite{cela2013quadratic}. Let's consider \(n\) facilities and \(n\) locations, where a flow matrix \(A = [a_{ik}]\) quantifies the amount of goods flow between facilities \(i\) and \(k\), and a distance matrix \(B = [b_{j\ell}]\) stores the distance between locations \(j\) and \(\ell\). The goal is to assign each facility to a location such that no two facilities are linked to the same location, while minimizing the total cost defined as the sum of goods flow and distances, i.e., \(\sum_{i,k} a_{ik} b_{\pi(i)\pi(k)}\) where \(\pi\) is a permutation mapping facilities to locations. QAP is an NP-hard problem that is widely used in facility layout, scheduling, and network design \cite{commander2005survey}. For the QUBO representation of the problem, each possible assignment of facility \(i\) to location \(j\) is associated with \(x_{ij} \in \{0,1\}\), indicating whether the assignment is made or not. Hence, there are $N=n^2$ binary variables. The goal is to minimize the total cost while maintaining a unique facility-to-location link.
The cost function to minimize (\(C_{\text{QAP}}(x)\)) is,
\begin{equation}
    C_{\text{QAP}}(x) = \sum_{i,k=1}^n \sum_{j,\ell=1}^n a_{ik} b_{j\ell} \, x_{ij} x_{k\ell} + P_1 \sum_{i=1}^n \left( \sum_{j=1}^n x_{ij} - 1 \right)^2 + P_2 \sum_{j=1}^n \left( \sum_{i=1}^n x_{ij} - 1 \right)^2,
    \label{qapC}
\end{equation}
where \(a_{ik}\) is the flow between facilities \(i\) and \(k\), \(b_{j\ell}\) is the distance between locations \(j\) and \(\ell\), and \(P_1, P_2 > 0\) are penalty coefficients. The first term is the quadratic assignment cost, while the second and third terms enforce the one-to-one facility-to-location constraint. To construct the corresponding Ising Hamiltonian, we first map each of the $N=n^2$ binary variable to a spin variable $\sigma_p \in \{-1, +1\}$ with the linear index $p = (i-1)n + j$ with $i,j \in \{1,\dots,n\}$ and $p \in \{1, \dots, N\}$. $i$ denotes the facility index of spin $p$ and $j$ is the location index. The problem is thus encoded as the following Hamiltonian,  
\begin{equation}
    \hat{H}_{QAP} = \sum_{p=1}^{N} h_p \, \hat{\sigma}^z_p + \sum_{p=1}^{N-1} \sum_{q=p+1}^{N} J_{pq} \, \hat{\sigma}^z_p \hat{\sigma}^z_q + C,
    \label{qapH}
\end{equation}
where the local fields \(h_{p}\) and couplings \(J_{pq}\) are shown below. 
$h_p$ for a spin $p$ corresponding to the assignment pair $(i,j)$,
\begin{equation}
h_p = -\frac{1}{2} \sum_{k=1}^n \sum_{\ell=1}^n a_{ik} \, b_{j\ell} - \frac{(n-2)}{2} (P_1 + P_2)
\end{equation}
This is based on the assumption that the flow matrix $A$ and distance matrix $B$ are symmetric. If they are not symmetric, the first term should be $-\frac{1}{4} \sum_{k,\ell} (a_{ik}b_{j\ell} + a_{ki}b_{\ell j})$.

There are multiple cases for the interaction strength between the spins $p \neq q$, with $p = (i,j)$ and $q = (k,\ell)$.
\begin{itemize}
    \item Different Facility and Different Location ($i \neq k$ and $j \neq \ell$): $J_{pq} = a_{ik} \, b_{j\ell}/2$
    \item Same Facility, Different Location ($i = k$ and $j \neq \ell$): $J_{pq} = P_1/2$
    \item Different Facility, Same Location ($i \neq k$ and $j = \ell$): $J_{pq} = P_2/2$.
    \item Same Facility and Same Location ($i = k$ and $j = \ell$): $J_{pq} = J_{pp} = 0$ .
\end{itemize}
The system naturally forms frustrated energy landscapes characteristic of NP-hard optimization \cite{codognet2022quantum}.

\subsection{Binary Clustering}

The \textbf{binary clustering} problem (also known as graph partitioning) aims to partition a dataset or graph into two parts such that dissimilarity within clusters is minimized and across clusters is maximized \cite{zhang2018binary}. Take a set of \(n\) data points, a dissimilarity matrix \(W = [w_{ij}]\) is given, which encodes the weight between data points \(i\) and \(j\). The objective is to define two disjoint clusters such that the total weight from the dissimilarity matrix connecting points in different clusters is maximized, i.e., \(\sum_{i \in C_1, j \in C_2} w_{ij}\) is maximized. Binary clustering is widely used in image segmentation, community detection, and unsupervised machine learning applications where grouping items is essential \cite{pothen1997graph,wang2023graph}. For binary representation, each data point is associated with \(x_i \in \{0,1\}\) indicating its cluster assignment, with the goal being to maximize the cut weight, i.e., points with large dissimilarity weights in different clusters.
The cost function to minimize (\(C_{\text{BC}}(x)\)) is thus given as,
\begin{equation}
    C_{\text{BC}}(x) = -\sum_{i<j} w_{ij} (x_i + x_j - 2x_i x_j),
    \label{bcC}
\end{equation}
where \(w_{ij} \geq 0\) is the \textbf{dissimilarity} weight between points \(i\) and \(j\). The term promotes points with large dissimilarity weights being assigned to different clusters. \(-(x_i + x_j - 2x_i x_j) = -1\) iff \(x_i \neq x_j\), hence it directly maximize the cut weight \(\sum_{i \in C_1, j \in C_2} w_{ij}\). The corresponding Ising Hamiltonian is,
\begin{equation}
    \hat{H}_{BC} = \sum_{i<j} \frac{w_{ij}}{2} \hat{\sigma}^z_i \hat{\sigma}^z_j + \text{const},
    \label{bcH}
\end{equation}
where the couplings are \(J_{ij} = \frac{w_{ij}}{2}\) (antiferromagnetic interactions), and the local fields \(h_i = 0\). This is the same structure as a Max-cut graph problem, which is NP-hard \cite{rodriguez2020clustering}. 

\subsection{Protein Folding (HP Model)}

The \textbf{protein folding} problem involves finding the spatial configuration of a chain of amino acids that minimizes its energy \cite{compiani2013computational}. In simplified lattice models such as the HP (hydrophobic-polar) model, each amino acid is classified as either hydrophobic (H) or polar (P), and the chain is constrained to a discrete grid where adjacent amino acids in the sequence are brought close together, mimicking real biological folding behavior \cite{stillinger1993toy}. The goal is to find a self-avoiding walk on the lattice that maximizes contacts between hydrophobic amino acids, as these contacts drive the folding process, which reduces free energy. 
Understanding protein folding is crucial in biology, medicine, and drug design, as a protein’s function depends on its structure \cite{dill2008protein}. Even simplified versions of the problem are computationally challenging (NP-hard). These toy models are often mapped to combinatorial optimization problems with constraints and pairwise interactions between hydrophobic residues that are adjacent on the lattice but not adjacent in the sequence \cite{fraenkel1993complexity}. We consider a minimal toy model that captures the tension between \textbf{hydrophobic attraction} and \textbf{spatial exclusion} in protein folding. The model consists of $L$ amino acids on a lattice. The goal is to minimize the free energy by maximizing favorable hydrophobic contacts while respecting the geometric constraints of the lattice.
The QUBO formulation and the cost function (\(C_{\text{PF}}(x)\)) for this simplified HP model with a protein sequence consisting of $L$ amino acids is,
\begin{equation}
    C_{\text{PF}}(x) = -\sum_{i<j} c_{ij} \, x_{ij} + P_1 \sum_{i<j} x_{ij} + P_2\sum_{((i,j),(k,\ell)) \in \mathcal{C}} x_{ij} x_{k\ell},
    \label{pfC_toy}
\end{equation}
where \(x_{ij} \in \{0,1\}\) indicates whether amino acids \(i\) and \(j\) form a non-bonded hydrophobic contact, \(c_{ij} = 1\) if both residues \(i,j\) are hydrophobic and non-adjacent in the sequence (with \(|i-j| > 1\)) (\(c_{ij} = 0\) otherwise), and $\mathcal{C}$ is the set of mutually exclusive contact pairs, generally precomputed based on the lattice geometry. The first term rewards favorable H-H contacts by lowering the energy of the selected pairs. The second term, with penalty \(P_1 > 0\), is a linear self-avoidance constraint between all contacts equally, and the third term is a quadratic penalty $P_2 > 0$ that enforces spatial exclusion geometrically. 
To represent this cost function as an Ising Hamiltonian, we define $N = \frac{L(L-1)}{2}$ binary variables corresponding to all pairs $(i,j)$ with $1 \le i < j \le L$.
Each pair $(i,j)$ is mapped to an index $p = (i-1)L - \frac{i(i+1)}{2} + j$ such that $p \in \{1, \dots, N\}$. 
The Ising Hamiltonian after substituting $x_{ij} = \frac{1 - \hat{\sigma}_p}{2}$ is thus,
\begin{equation}
\hat{H}_{PF} = \sum_{p=1}^{N} h_p \, \hat{\sigma}^z_p + \sum_{p=1}^{M-1} \sum_{q=p+1}^{N} J_{pq} \, \hat{\sigma}^z_p \hat{\sigma}^z_q + \text{const},
\label{pfH_toy}
\end{equation}
where the longitudinal field $h_p = -(P_1 - c_p)/2 - \sum_{q: (p,q) \in \mathcal{C}} P_2/4$, the interaction $J_{pq} = P_2/4$ if $(p,q)\in \mathcal{C}$ and $J_{pq}=0$, otherwise. This simplified model captures the essential tension between hydrophobic attraction and spatial exclusion that drives protein folding, while remaining small enough to serve as a pedagogical toy problem.

\section{LOQAL: Localized Optimal Control for Quantum Annealing in a Loop}
\label{LOQAL}
In the previous section, multiple optimization problems and the corresponding Ising Hamiltonians encoding their solutions as the ground state are presented. By extending the work in Ref.~\refcite{kapil}, we outline the steps to simulate the dynamics of these Hamiltonians using Rydberg atoms and find the ground state/s. A non-blockade-based encoding scheme using local detuning is shown as a single framework in Sec.~\ref{LD}, mapping the Ising Hamiltonians (Eqs.~\ref{msatexH},\ref{setpackH},\ref{qapH},\ref{bcH},\ref{pfH_toy}) to the controllable Rydberg Hamiltonian (Eq.~\ref{eq:IsingHam}). The ground state of the target problem Hamiltonian is dynamically reached using an optimal quantum annealing method, as discussed in Sec.~\ref{OCAL}.

\subsection{Local detuning encoding}
\label{LD}
A general method for encoding any problem into a Rydberg system is first to represent the problem cost function in the spin Hamiltonian $H_G$,
\begin{equation}
    \hat{H}_{G} = \sum_{i} h_{i} \hat{\sigma}^z_{i} + \sum_{(i,j)} J_{ij} \hat{\sigma}^z_{i} \hat{\sigma}^z_{j},
    \label{GH}
\end{equation}
where $h_i$ is the longitudinal field and $J_{ij}$ are the spin interaction terms. By comparing the spin representation of the Rydberg Hamiltonian (Eq.~\ref{eq:IsingHam}) and the spin Hamiltonians of the cost functions (Eq.~\ref{GH}), we define the detunings $\Delta_j$ in a manner such that the effective longitudinal field $h_i$ is fine-tuned to simulate the problem Hamiltonian. Similarly, the distances between the atoms are controlled to fix the interaction strength $V_{jk}$ between the atoms $j,k$, which corresponds to the spin interaction $J_{jk}$.
Thus, the general mapping is,
\begin{align}
V_{ij} = 4J_{ij}
\end{align}
\begin{align}
\Delta_j =  2h_j - 2 \sum_{k \neq j}J_{jk} = 2h_j - \frac{1}{2} \sum_{k \neq j} V_{jk}
\label{LD_all}
\end{align}
This mapping also implies that the value of the detunings and the interactions are related to the specific problem structure. We provide the corresponding encoding below for different problems.
\begin{romanlist}
    \item Mixed two-SAT and XOR-SAT problem: For this problem, an example is taken to demonstrate the encoding scheme, as the specific values of the longitudinal fields and the spin interaction can vary vastly depending on the problem constraints. However, the key step is to add penalty terms for the constraints in both problems (two-SAT and XOR-SAT) and substitute the binary variables with Pauli matrices to get the spin Hamiltonian. The detuning and the interaction term for the example problem are calculated by comparing Eq.~\ref{msatexH} and Eq.~\ref{eq:IsingHam} as,
    \begin{align}
    V_{ij} = \frac{1}{2}+\frac{P}{4} ,\ V_{jk} = \frac{P}{4} \nonumber
    \end{align}
\begin{align}
\Delta_i = \Delta_k = \frac{P}{4} ,\ \Delta_j = \frac{P}{2}.
\label{LDXSAT}
\end{align}

    \item Set packing: The weighted set packing problem is chosen, with weights $w_i$ for a subset $S_i$, $D$ being the set of conflicting pairs, $d_i$ is the number of subsets overlapping with $S_i$, and $P$ is the penalty. From Eq.~\ref{setpackH} and Eq.~\ref{eq:IsingHam} 
        \begin{align}
    V_{ij} = \frac{P}{4} ,\ \text{for } (i,j)\in D \nonumber
    \end{align}
\begin{align}
\Delta_i = \frac{w_i}{2} - \frac{d_iP}{4}.
\label{LDSP}
\end{align}
    \item QAP: A general problem and the cost function is defined with $n$ locations and $n$ facilities. The parameters $p=(i,j)$, $q=(k,l)$, where $i,k$ are the variables representing facilities and $j,l$ are the location variables. From Eq.~\ref{qapH} and Eq.~\ref{eq:IsingHam}, 
\begin{align}
    \text{(i) }V_{pq} = 2a_{ik}b_{jl}\ (\text{ for } i\neq k,j \neq l) \nonumber \\ \text{(ii) }V_{pq} = 0 \ (\text{for } i= k,j= l)\nonumber \\
    \text{(iii) }V_{pq} = 2P_1 \ (\text{for } i= k,j \neq l) \nonumber \\ \text{(iv) }V_{pq} = 2P_2 \ (\text{for } i\neq k,j= l)\nonumber 
\end{align}
\begin{align}
\Delta_p = -\frac{1}{2} \sum_{k=1}^n \sum_{\ell=1}^n a_{ik} \, b_{j\ell} - \frac{(n-2)}{2} (P_1 + P_2) -  \frac{1}{2} \sum_{p \neq q} V_{pq}.
\label{LDQAP}
\end{align}

    \item Binary clustering: A weighted problem is defined with $n$ data points and a dissimilarity matrix $W=[w_{ij}]$ encoding the weights between the points $i$ and $j$. The problem maps to a weighted max-cut problem, hence the encoding from Eq.~\ref{bcH} and Eq.~\ref{eq:IsingHam} is,
    \begin{align}
    V_{ij} = 2w_{ij} \nonumber 
    \end{align}
\begin{align}
\Delta_j =  - \sum_{k \neq j}^{n} w_{jk}.
\label{LDBC}
\end{align}
    
    \item Toy protein folding: The toy model consisting of $L$ amino acids capturing the hydrophobic attraction, the tension in the chain due to geometry and spatial exclusion, through the coefficients $c_{ij}$ and the penalties $P_1,P_2$ respectively. For a chain of length $L$, $N=L(L-1)/2$ binary variables are required to define a spin Hamiltonian; hence, a linear parameter $p$ is mapped to a pair $(i,j)$ to facilitate the construction of the Hamiltonian. From Eq.~\ref{pfH_toy} and Eq.~\ref{eq:IsingHam},
    \begin{align}
    V_{pq} = P_2\ (\text{if } p,q\in \mathcal{C}),\ V_{pq} = 0\ (\text{otherwise}) \nonumber
\end{align}
\begin{align}
\Delta_p = -(P_1 - c_p) - \sum_{q: (p,q) \in \mathcal{C}} \frac{P_2}{2} -  \frac{1}{2} \sum_{p \neq q} V_{pq}.
\label{LDPF}
\end{align}
\end{romanlist}

\noindent After mapping the problems to the corresponding target Rydberg Hamiltonians, they are solved by temporally varying the laser parameters to reach the specific choice of detuning values at the end of the protocol, while minimizing the energy of the full system. However, there is a caveat to this scheme due to the geometry restrictions of the physical system. The detunings are controlled via laser parameters, and the values can reach the limits that are physically difficult to achieve due to experimental restrictions. This can be remedied by rescaling the problem parameters and bringing the detuning values within the acceptable/achievable limits. On the other hand, due to distance-dependent interactions, the initial arrangement of the atoms is crucial for the scheme before any evolution of the many-body Hamiltonian. Since the problem parameters are related to the interactions $V(r_{ij})$, which in turn are reflected in the choice of the relative distances $r_{ij}$. The atoms are arranged in a manner that provides a true representation of problem interactions; however, \textit{unwanted interactions} can always become a serious issue for densely connected problems. Apart from unwanted interactions, the arrangement of atoms is also limited by the geometry. This case appears when the problem parameters are such that the resulting relative distances between the atoms cannot be realized on a 2D plane. However, we emphasize that this scheme remains applicable to a large class of QUBO problems, as shown in this work. Adding a third dimension can also increase the flexibility of the atom arrangement. 

\subsection{Optimal Quantum Annealing in a Loop with Rydberg atoms}
\label{OCAL}
The previous section outlined the encoding of the optimization problems onto the Rydberg spin Hamiltonian and discussed the caveats regarding the spatial arrangement of the atoms. The objective is to find accurate solutions to the problems efficiently with respect to the number of iterations and run-time. This is achieved by numerically solving the spin dynamics using the following Rydberg Hamiltonian for a specific arrangement of atoms,
\begin{align}
\hat{H}_{Ryd} = \frac{\Omega(t)}{2} \sum_{j} \hat{\sigma}^{x}_{j} - \sum_{j} \Delta_j(t) \hat{n}^{e}_{j}
+ \sum_{k,j} V_{kj}\hat{n}^{e}_{k}\hat{n}^{e}_{j} ,
\label{Ryd1} 
\end{align}
where $\hat{\sigma}_{j}^x = \ket{e}_j \bra{g} +\ket{g}_j \bra{e}$  and excitation number operator $\hat{n}^{e}_{j} = \frac{1}{2} \qty(\hat{\sigma}^{z}_{j} + \mathbb{I} )$ as $\ket{e}_j \bra{e}$. The objective is to reach the ground state of the target Hamiltonians $\hat{H}_{target}$ given by Eqs.~\ref{msatexH},\ref{setpackH},\ref{qapH},\ref{bcH},\ref{pfH_toy}. This is done by optimizing the laser parameters to minimize the energy of the target Hamiltonian with respect to the instantaneous ground state. The parameters are defined so that after an evolution time $T$, the Rydberg Hamiltonian transforms into the problem Hamiltonian as defined by the encoding scheme. \\

\noindent \textit{Detuning}:
All atoms are initialized in a relatively easy-to-prepare state, e.g., all in the ground state $\ket{gg...g}$, which corresponds to a large non-zero detuning value $\Delta_j(t=0) \neq 0$ or, e.g., all in the excited state $\ket{ee...e}$ leading to a large negative detuning value. 
Next, the time-dependent detuning of each atom, $\Delta_{j}(t)$, is varied to reach a specific value at the end of the protocol as defined by Eqs. \ref{LDXSAT}, \ref{LDSP}, \ref{LDQAP}, \ref{LDBC}, and \ref{LDPF}. A single time-dependent parameter $\Delta_G(t)$ governs all detunings in the problem, which is then optimized. 
Let $C = [\Delta_1(T), \Delta_2(T), \dots, \Delta_N(T)]$ be the predefined final detunings for $N$ atoms, with fixed relative ratios. $\Delta_G(t)$ is then given as, 
\[
\Delta_j(t) = \Delta_G(t)\,\Delta_j(T).
\]
The function $\Delta_G(t)$ starts negative at $t=0$ and increases to $1$ at $t=T$, reaching the desired final detuning values. \\

\noindent \textit{Rabi frequency}:
The boundary conditions of the Rabi frequency are such that the initial and final values are zero. At intermediate times, non-zero values of the Rabi frequency $\Omega(t)$ provide a transverse field for the above Ising Hamiltonian. 
This leads to dynamical coupling of different many-body states, thereby accessing a larger part of the Hilbert space. This is key to the quantum annealing process as it explores different configurations by applying a non-zero transverse field while traversing the energy landscape to get to the desired many-body ground state configuration. \\

The idea is to use the techniques of optimal control theory to reach this state efficiently, well within the system lifetime \cite{rabitz, kelly2014optimal,li2017hybrid,CRAB1, PhysRevA.91.052306, Mukherjee_2020, Mukherjee2}. In particular, a combination of gradient and non-gradient-based methods are used to shape the pulses in time which allows us to steer the many-body state optimally towards the true solution \cite{polak2012optimization,hasdorff1976gradient,conn2009introduction,hare2013survey,broyden1970convergence,fletcher1970new,goldfarb1970family,shanno1970conditioning,nelder1965simplex}. Specifically, the hybrid optimization approach involves sandwiching a non-gradient based optimizer in between two gradient-based ones. Intuitively, this allows the system to reach low energy subspaces with a gradient-based optimizer, navigate local minimas using a non-gradient-based optimizer and fine-tuning towards the global minima using the last gradient-based optimizer. The detuning and the Rabi frequency pulses are constructed using Fourier basis in some cases and splines in other cases to get smoother pulses. The objective function that needs to be minimized during the optimal control is the expectation value of the instantaneous many-body state $\ket{\psi_{inst}(t)}$ with respect to the target Hamiltonian $\hat{H}_{target}$, which is given as 
\begin{align}
E = \bra{\psi_{inst}(t)}\hat{H}_{target} \ket{\psi_{inst}(t)} .
\label{expect}
\end{align}
$E$ is a useful observable as it can be measured experimentally. Apart from $E$, we define the overlap between the instantaneous many-body state and the many-body ground state $\ket{\psi_{g}}$ of the problem Hamiltonian $\hat{H}_{target}$ during the protocol given as
\begin{align}
F(t) = \sum_{g} |\braket{\psi_{inst}(t)}{\psi_{g}}|^2 .
\label{fid}
\end{align}
The fidelity $F$ is calculated over all the degenerate ground states. Furthermore, we use the approximation ratio to quantify the algorithm's performance. It shows how close the solution provided by an algorithm is to the true solution for a given problem. To benchmark quantum algorithms, the approximation ratio is defined numerically as follows, \cite{vazirani2001approximation}, 
\begin{align}
R= \frac{C_{max}-C_{obt}}{C_{max}-C_{opt}} ,
\label{Approxratio}
\end{align}
where $C_{opt}$ is the optimal value of the cost function, $C_{max}$ is the maximum value of the cost function, and $C_{obt}$ is the obtained value of the cost function evaluated through our method.

\section{Hardness of the optimization problems}
\label{HP}

There are many factors, such as problem structure, optimization landscape, number of optimal solutions, etc., that determine whether a problem is difficult or easy to solve using any given method. Taking inspiration from the physical system while accounting for all of these factors, we define a \textit{method-independent hardness parameter} $\mathcal{HP}_{MI}$ for any QUBO problem, extending the one defined in Ref.~\refcite{kapil}. 

Consider a spin Hamiltonian $H_{T}$ corresponding to a given QUBO problem, encoding it to $N$ qubits. Let the spectral decomposition of this Hamiltonian be defined as $H_T = \sum_{k=0}^{2^N-1} E_k \ketbra{E_k}{E_k}$ with eigenvalues $E_k$, ordered as $E_0 \leq E_1 \leq \cdots \leq E_{2^N-1}$, and eigenstates $\ket{E_k}$. These eigenstates, in general, will form degenerate or near-degenerate subspaces, all of which, except the ground state subspace, contribute to the difficulty of a particular problem. The degeneracy for each such subspace $\alpha$ is defined as $D_\alpha = |\{E_k : |E_k - \bar{E}_\alpha| < \epsilon\}|$ states, with $\epsilon$ being the tolerance for near-degeneracy. The optimal (ground) subspace has energy $E_{\text{opt}} = \bar{E}_0$ and degeneracy $D_{\text{opt}} = D_0$. All the excited subspaces other than the ground state are termed as suboptimal.
The subspace closest to the ground state/subspace in terms of energy defines a fundamental scale in the problem, measuring how well the optimal and the suboptimal configurations are separated. This is termed as the spectral gap $G_\Delta = \bar{E}_1 - \bar{E}_0$. 
While finding the ground state of two Ising systems, the one with a larger gap is relatively easier to solve compared to the one with a lower gap value. Apart from the first excited subspace and its energy gap, further excited subspace can also be a threat to finding the optimal solution, although not as much. We categorize a subspace $\alpha > 0$ to be threatening if it correspond to the first excitation state, $\bar{E}_1$, or if it has high degeneracy, $D_\alpha \geq D_{\text{thresh}}$ where $D_{\text{thresh}} = \max(1, D_{\text{opt}}/2)$ is the threshold we used for filtering the subspaces. We define $\Sigma$ that encapsulates the hardness due to the degenerate subspaces as, 
\begin{equation}
\Sigma = \sum_{\alpha \in \mathcal{T}} D_\alpha \, \exp\pqty{-\frac{\bar{E}_\alpha - \bar{E}_0}{G_\Delta}}.
\end{equation}
where the exponential factor ensures that subspaces far in energy contribute less, which are suppressed exponentially with their energy gap to the ground state.

Thus, the hardness parameter combining the degenerate subspace structure along with the energy gap of the suboptimal configurations with the ground state is defined as,
\begin{align}
\mathcal{HP}_{MI} = \frac{\Sigma}{|E_{\text{opt}}| \cdot D_{\text{opt}}} \cdot \frac{1}{G_\Delta^2} \nonumber
\end{align}
\begin{align}
\mathcal{HP}_{MI} = \frac{1}{|E_{\text{opt}}| \cdot D_{\text{opt}} \cdot G_{\Delta}^2} 
\sum_{\substack{\alpha > 0 \\ \bar{E}_\alpha - \bar{E}_0 \leq G_\Delta \\ \text{or } D_\alpha \geq D_{\text{thresh}}}}
D_\alpha \, \exp\pqty{-\frac{\bar{E}_\alpha - \bar{E}_0}{G_\Delta}}.
\end{align}
The factor $1/|E_{\text{opt}}|$ is used for normalizing the energy scale of the problem, which makes sure that the energy differences are measured relative to the ground state energy. $\Sigma/D_{\text{opt}}$ is the \textit{relative degeneracy factor}. The problem becomes harder to solve when the degeneracy of suboptimal subspaces is comparable to that of the optimal subspace, as these states act as a trap for the population. $1/G_\Delta^2$ is the gap contribution to the hardness; a small gap leads to densely packed states near the optimal subspace, making it difficult to isolate the ground state.
\begin{figure}[t]
\centerline{\includegraphics[width = 1\linewidth,trim={1cm 6cm 1cm 10cm},clip]{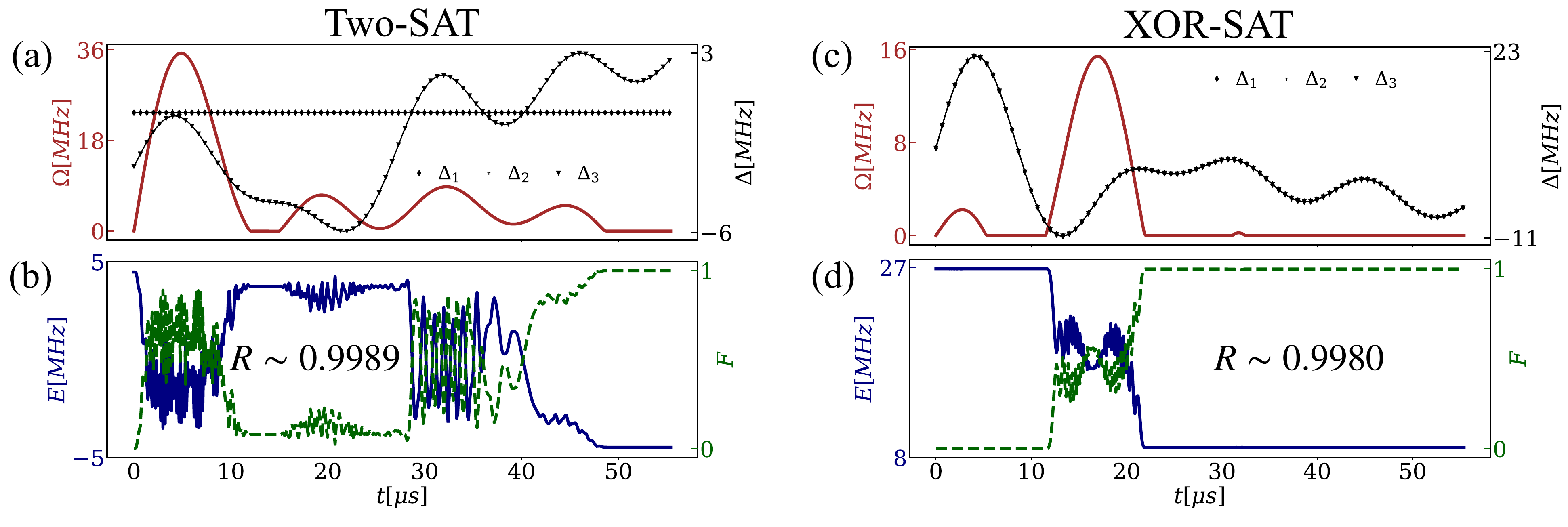}}
\caption{Solutions to the two-SAT [panels (a)-(b)] and XOR-SAT problems [panels (c)-(d)] as defined by Eqs.~\ref{satex}~and~\ref{xorex} found using the optimal quantum annealing method. [(a),(c)] shows the optimal protocols for the Rabi frequency $\Omega(t)$ in solid dark red curve as well as the detunings $\Delta_i(t)$ in dotted black symbols with time $t$ for the two-SAT and XOR-SAT respectively. The expectation value $E$ of the final Hamiltonian with respect to the instantaneous state is minimized, depicted as a solid blue curve in [(b)] for two-SAT and [(d)] for XOR-SAT. Panels [(b),(d)] also show the fidelity $F$ in a dashed green curve along with the final approximation ratio $R$ in text for the respective problems.}
\label{f1}
\end{figure}
This is a method-independent parameter, as only the spectral properties of the problem Hamiltonian are enough to define it. The different parts of the parameter can be used to define hardness for a particular method, for example, adiabatic methods fail when $G_\Delta$ is small because the required evolution time scales as $1/G_\Delta^2$. While the optimal control methods face issues due to $\Sigma$ being large, as many transitions must be suppressed, and variational methods encounter barren plateaus when many near-degenerate subspaces exist (a large number of $\alpha>0$ subspaces close to the ground state).

\section{Results}
\label{Results}
In this section, we solve prototypical optimization problems numerically using the encoding and the optimal quantum annealing method as discussed in the previous sections. We also find the hardness parameter of the problems using corresponding Ising Hamiltonians to analyze the results and compare the problems. 
The physical parameters considered for the numerical study are: Rydberg states $60S_{1/2}$ of the Cs atoms, van der Waals coefficient $C_6\sim 139~GHz \cdot \mu m^6$, and a radiative lifetime of $\sim 234 \mu s$ \cite{Gal,feng2009lifetime,vsibalic2017arc}

\subsection{Optimal protocol for individual problem instances}

For each problem, we consider a specific example and minimize the expectation value of the instantaneous ground state with respect to the corresponding target Hamiltonian. We address relatively small sized example problems only for the proof of principle of our scheme. Fidelity of the final state is also calculated as a post processing step to validate the optimality of the obtained state. This require the ground state exact configuration apriori, which is relatively easier for small sized problems.
In the corresponding figures for all the problems, the top panels show the optimal profiles for the Rabi frequency \(\Omega\) and the detunings \(\Delta_i\) over the protocol time \(t\). The lower panels illustrate the corresponding energy minimization \(E\) for the given profile. The fidelity \(F\) of the solution is plotted in the lower panel to verify whether the ground state has been reached, along with the final approximation ratio \(R\) of the solution.

\begin{figure}[t]
\centerline{\includegraphics[width = 0.8\linewidth,trim={10cm 6cm 11cm 4cm},clip]{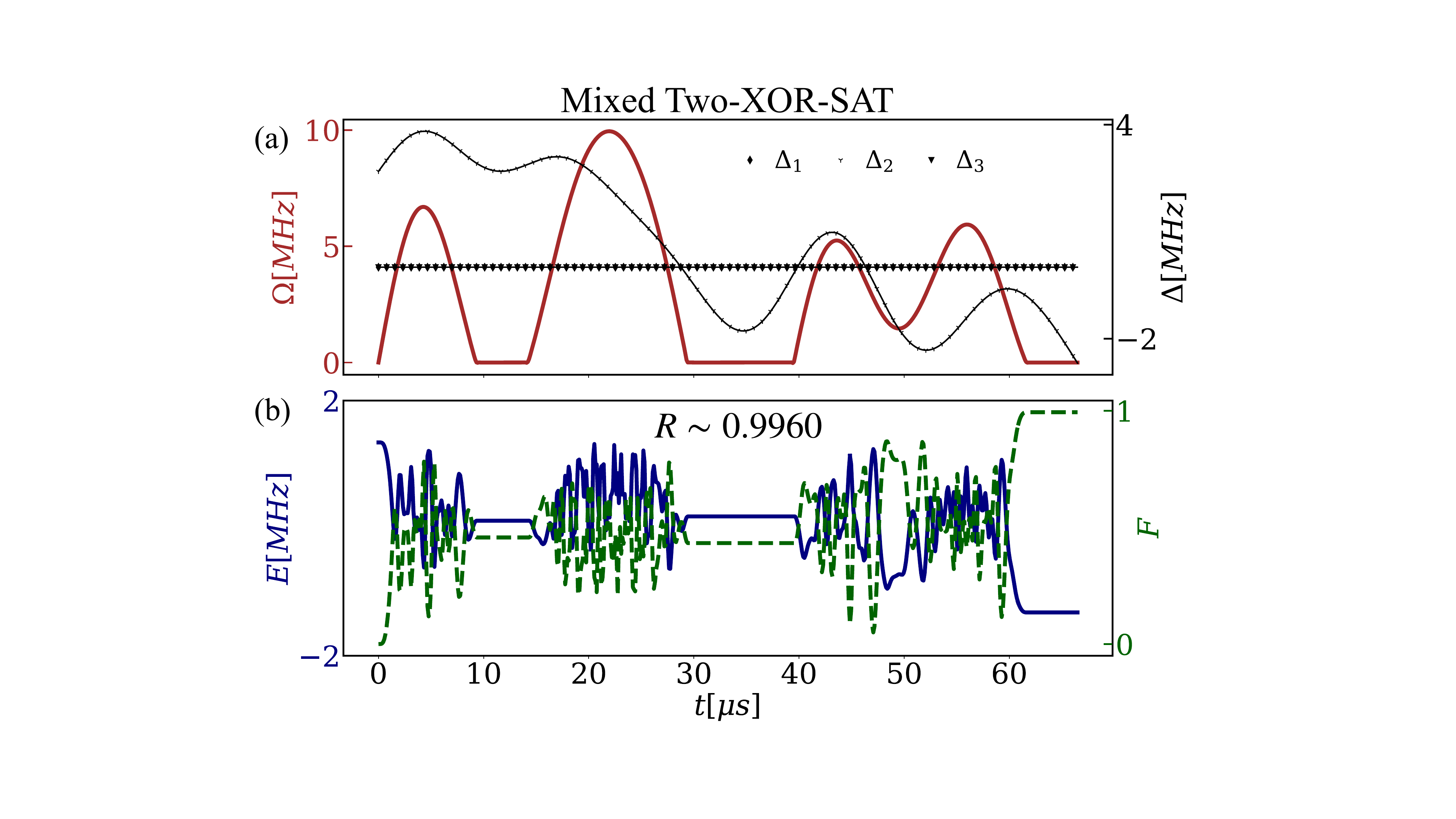}}
\caption{
The figure shows the optimal protocol for finding the solution of the mixed-two-XOR-SAT problem, given by Eq.~\ref{mxsatex}. The panel [(a)] shows the optimal protocols for the Rabi frequency $\Omega(t)$ in a solid dark red curve and the detunings $\Delta_i(t)$ in dotted black symbols with time $t$. In panel [(b)], the expectation value $E$ (solid blue curve) that is minimized by the optimizer is shown along the fidelity $F$ (dashed green curve) and the approximation ratio $R$ (text).
}
\label{f2}
\end{figure}

In Fig.~\ref{f1}, simple examples of the two-SAT (panels (a) and (b)) and XOR-SAT (panels (c) and (d)) problems are chosen to be solved. For the Two-SAT problem, the constraints are as follows,
\begin{align}
   (x_1 \lor x_2) \land (\neg x_1 \lor x_3) 
   \label{satex}
\end{align}
which is equivalent to $\neg x_1 \Rightarrow x_2, x_1 \Rightarrow x_3$.
The problem is satisfiable with multiple solutions.
For example, if \(x_1 = 0\), then \(x_2 = 1\) and \(x_3 = 1\); if \(x_1 = 1\),
then \(x_3 = 1\) while \(x_2\) can take any values. The problem is tractable as there are no cycles in the system, leading to a frustration-free optimization landscape. The fidelity and the approximation ratio, as shown in Fig.~\ref{f1}b, reach very close to the optimal values ($\sim 0.999$). The corresponding optimal protocol shown in panel (a) uses the Fourier basis for its construction. The problem is relatively easier to tackle as numerically multiple such protocols were found using other bases (e.g., splines) as well as other initializations, leading to the conclusion that there is an abundance of optimal protocols for this problem. Another key point is that out of three detuning values for each variable, only one is non-zero, indicating that the dynamics of a single atom drives the system towards the solution rather than competition between all three. 
\begin{figure}[t]
\centerline{\includegraphics[width = 0.8\linewidth,trim={10cm 6cm 11cm 4cm},clip]{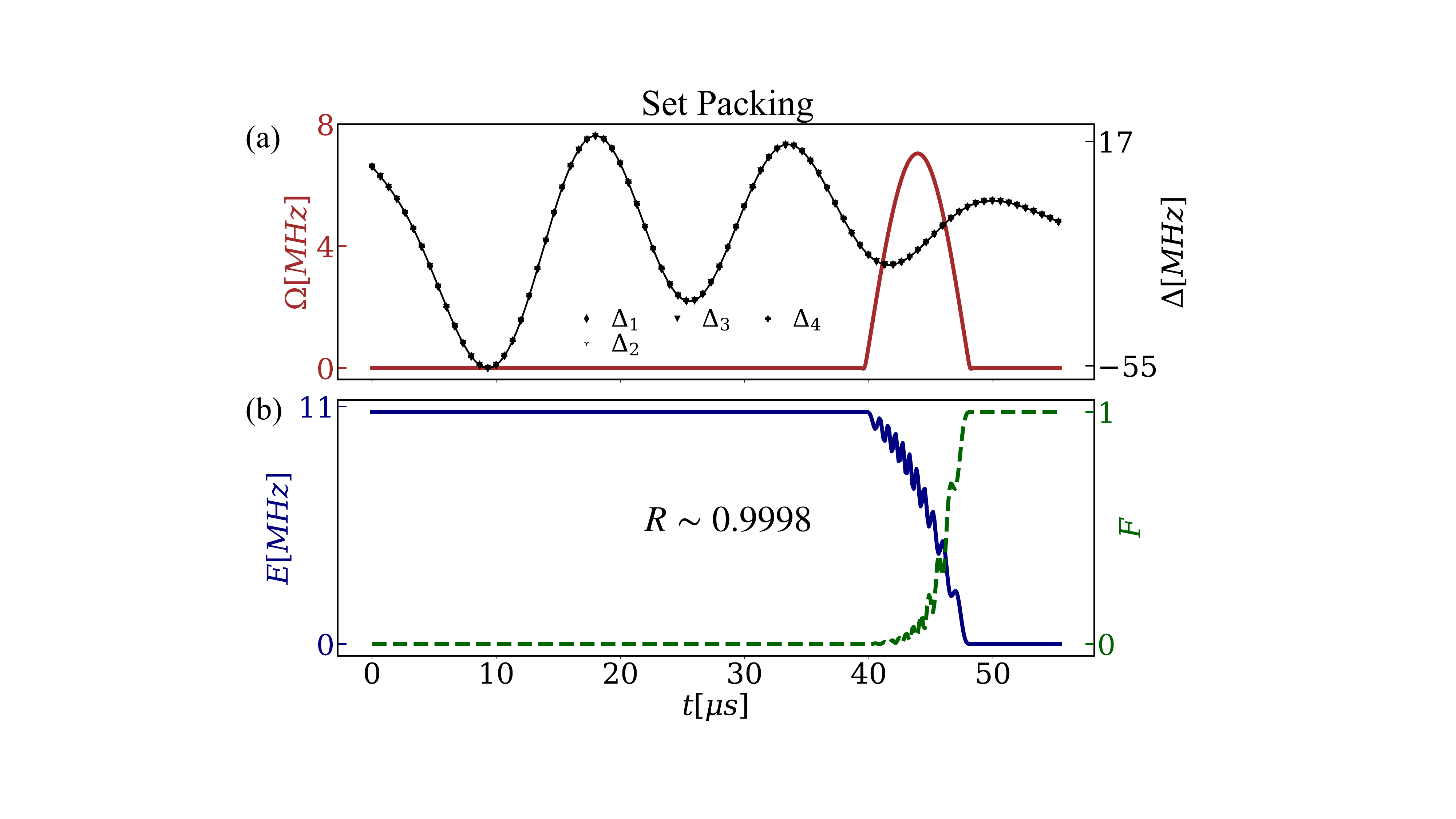}}
\caption{Optimal protocol for solving the set packing problem (Eq.~\ref{spex}) where the ground state of the corresponding Hamiltonian is found. [(a)] contains the optimal Rabi frequency (solid dark red), optimal detunings (dotted markers), and [(b)] shows the corresponding expectation value $E$ (solid blue), the fidelity $F$ (dashed green), and the approximation ratio $R$ (in text).}
\label{f4}
\end{figure}
A similar behavior is observed for the XOR-SAT problem, for which the example problem is defined as follows,
\begin{equation}
    x_1 \oplus x_2 = 1, \qquad
x_2 \oplus x_3 = 1, \qquad
x_3 \oplus x_1 = 1.
\label{xorex}
\end{equation}
This problem instance contains a frustrated cycle, an odd cycle of parity constraints that cannot be satisfied simultaneously. It is an example in which the inconsistency arises from global structure rather than local constraints. Although the problem is unsatisfiable, the Rydberg annealer finds the best assignment violating the minimum number of constraints with high fidelity as depicted in Fig~\ref{f1}d corresponding to the protocol shown in panel (c). The detuning value for each variable is the same due to the cyclic structure; each variable contributes similarly to the problem.

For both plots, the fluctuations in the energy $E$ (as well as fidelity $F$) increase during the protocol when the Rabi frequency is relatively large. This is because larger Rabi frequency leads to denser packing of energy gaps, promoting more non-adiabatic transitions in the system and hence, population jumps between many energy levels. Such transitions make the protocol time scales lower than that of adiabatic annealing and introduce quantum fluctuations to steer the system towards the solution using constructive interference.

The previous two problems, two-SAT and two-XOR-SAT, can be solved in polynomial time using classical algorithms; however, a combined problem (containing two-SAT and XOR-SAT clauses) becomes NP-hard. The combined problem for the numerical study is defined as,     
\begin{align}
   (x_1 \lor x_2) \land (\neg x_1 \lor x_3), \qquad
x_2 \oplus x_3 = 1. 
\label{mxsatex}
\end{align}
Fig~\ref{f2} shows the optimal protocols in panel (a) and the corresponding energy minimization in panel (b). The problem is considered difficult to solve because fewer optimal protocols were identified during the numerical study compared to the individual two-SAT and XOR-SAT problems. The final fidelity suddenly increases after the two overlapping Rabi frequency peaks between $t=40\mu s$ and $t = 60\mu s$ in panel (a). This represents a more sophisticated protocol than those shown in Fig.~\ref{f1}(a,c).

\begin{figure}[t]
\centerline{\includegraphics[width = 0.8\linewidth,trim={10cm 6cm 11cm 3cm},clip]{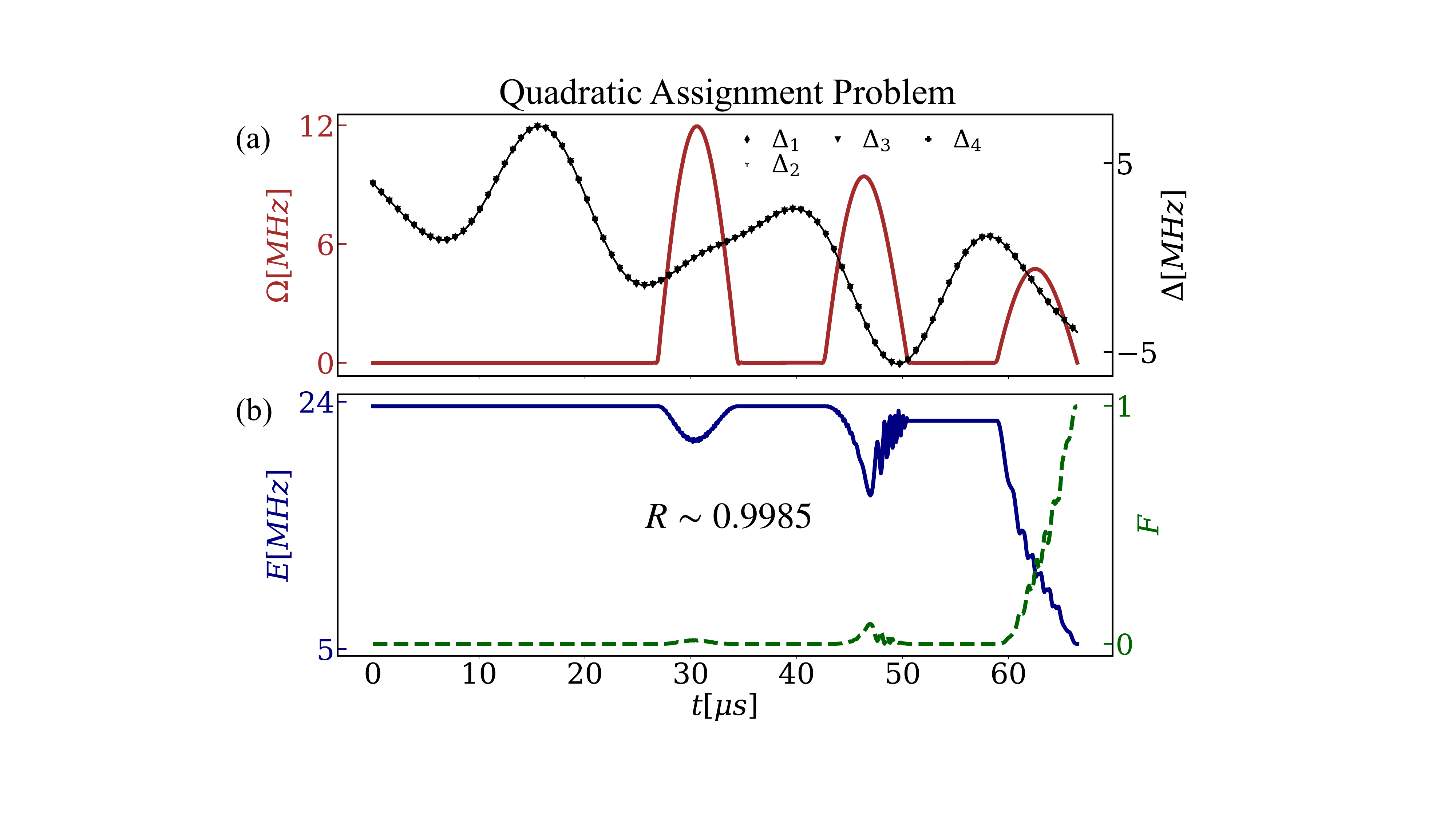}}
\caption{The quadratic assignment problem as defined by the matrices given in Eq.~\ref{qapex} is solved. The optimal Rabi frequency (solid dark red) and the optimal detunings (dotted markers) are plotted in [(a)]. The corresponding expectation value $E$ (solid blue), the fidelity $F$ (dashed green), and the approximation ratio $R$ (in text) are shown in [(b)].}
\label{f3}
\end{figure}

Consider a set packing problem with four sets and pairwise conflicts determined by overlaps. The problem is defined using binary variables \(x_i \in \{0,1\}\) as,
\begin{align}
    \max \sum_{i=1}^4 x_i \nonumber
\end{align}
\begin{align}
\text{subject to} \quad 
    x_1 + x_3 \le 1, \quad
x_1 + x_4 \le 1, \quad
x_2 + x_3 \le 1, \quad
x_2 + x_4 \le 1.
\label{spex}
\end{align}
A bipartite conflict graph can be constructed with partitions \(\{1,2\}\) and \(\{3,4\}\), and edges only across partitions. The Fig~\ref{f4} shows the optimal protocols in panel (a) and the energy (and fidelity) in panel (b). The problem is solved with very high accuracy, and many different protocols are possible to provide an optimal solution.
Similarly, in Fig~\ref{f3}, a small quadratic assignment problem is solved; in principle,  even a small instance of this problem is relatively difficult to solve. The problem we consider has a $2 \times 2 $ flow matrix and $2 \times 2 $ distance matrix,
\begin{align}
    F =
\begin{pmatrix}
0 & 3 \\
3 & 0
\end{pmatrix}, \qquad
D =
\begin{pmatrix}
0 & 2 \\
2 & 0
\end{pmatrix}.
\label{qapex}
\end{align}
where the binary variables \(x_{ik} \in \{0,1\}\) indicate whether facility \(i\) is
assigned to location \(k\), subject to assignment constraints $\sum_k x_{ik} = 1,\sum_i x_{ik} = 1$ with the objective $\min \sum_{i,j,k,l} F_{ij} D_{kl} x_{ik} x_{jl}$. The instance is symmetric and involves only two facilities and
locations, there are only two feasible assignments, both with the same
objective value. This makes the instance degenerate, with multiple global
optima. This problem becomes difficult to solve as encoding it as a QUBO requires a quadratic number of variables relative to the problem size. Since this problem is not native for QUBO encoding, the fidelity as shown in Fig.~\ref{f3}(b) is less than $1$, and this optimal protocol required a larger number of optimizer iterations as compared to all the other problems discussed so far.   

\begin{figure}[t]
\centerline{\includegraphics[width = 0.8\linewidth,trim={10cm 6cm 11cm 4cm},clip]{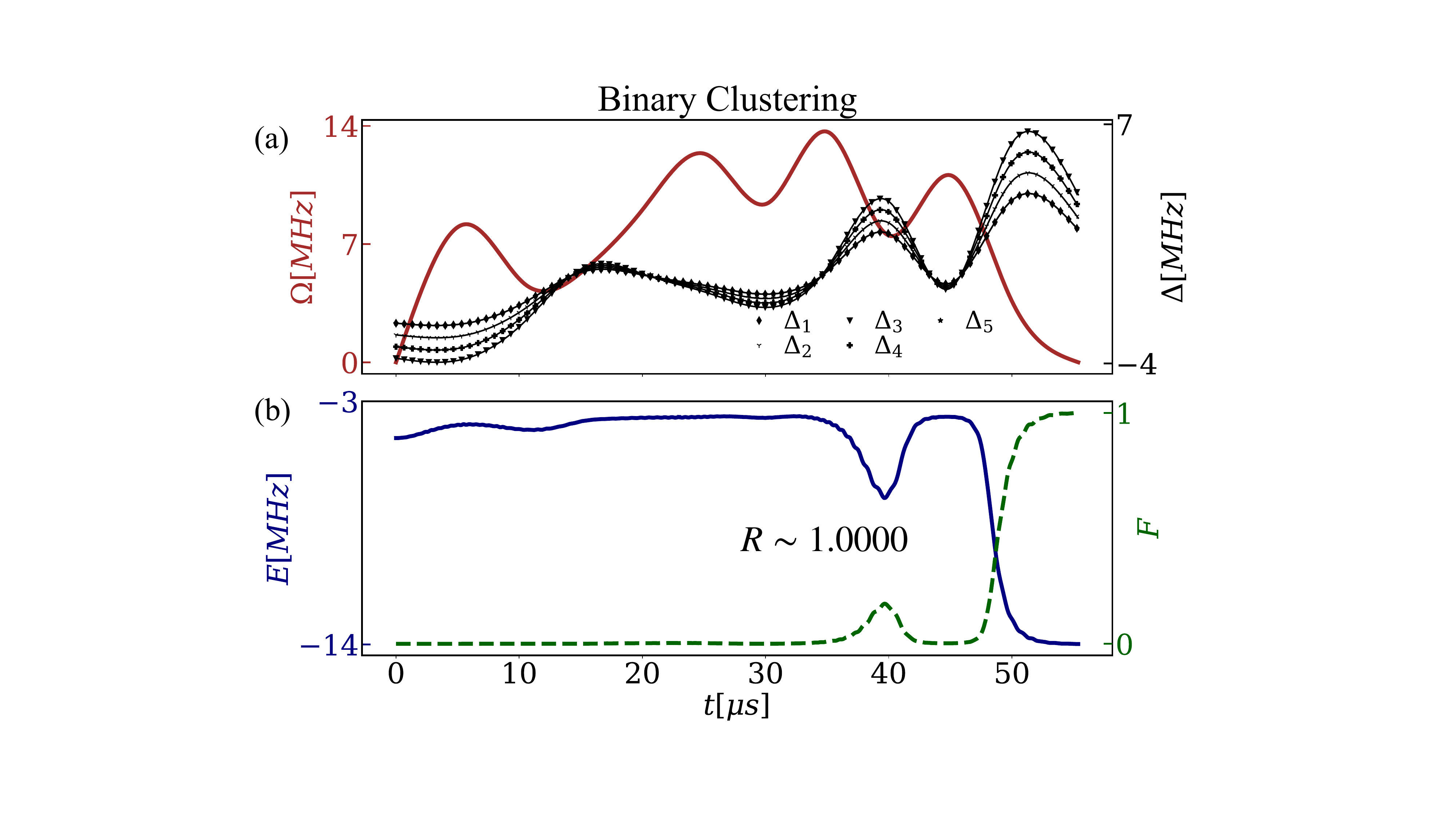}}
\caption{The figure shows the optimal protocol for finding the solution of a binary clustering problem (Eq.~\ref{bcex}). [(a)] the optimal protocols for the Rabi frequency $\Omega(t)$ in solid dark red curve and four different detunings $\Delta_i(t)$ in dotted black symbols with time $t$. [(b)] The expectation value $E$ (solid blue curve) is minimized, the fidelity $F$ (dashed green curve) for a sanity check of the obtained result, and the approximation ratio $R \sim 1$ (text) shows that the system reaches the optimal solution.}
\label{f6}
\end{figure}

\begin{figure}[t]
\centerline{\includegraphics[width = 0.8\linewidth,trim={10cm 6cm 11cm 3cm},clip]{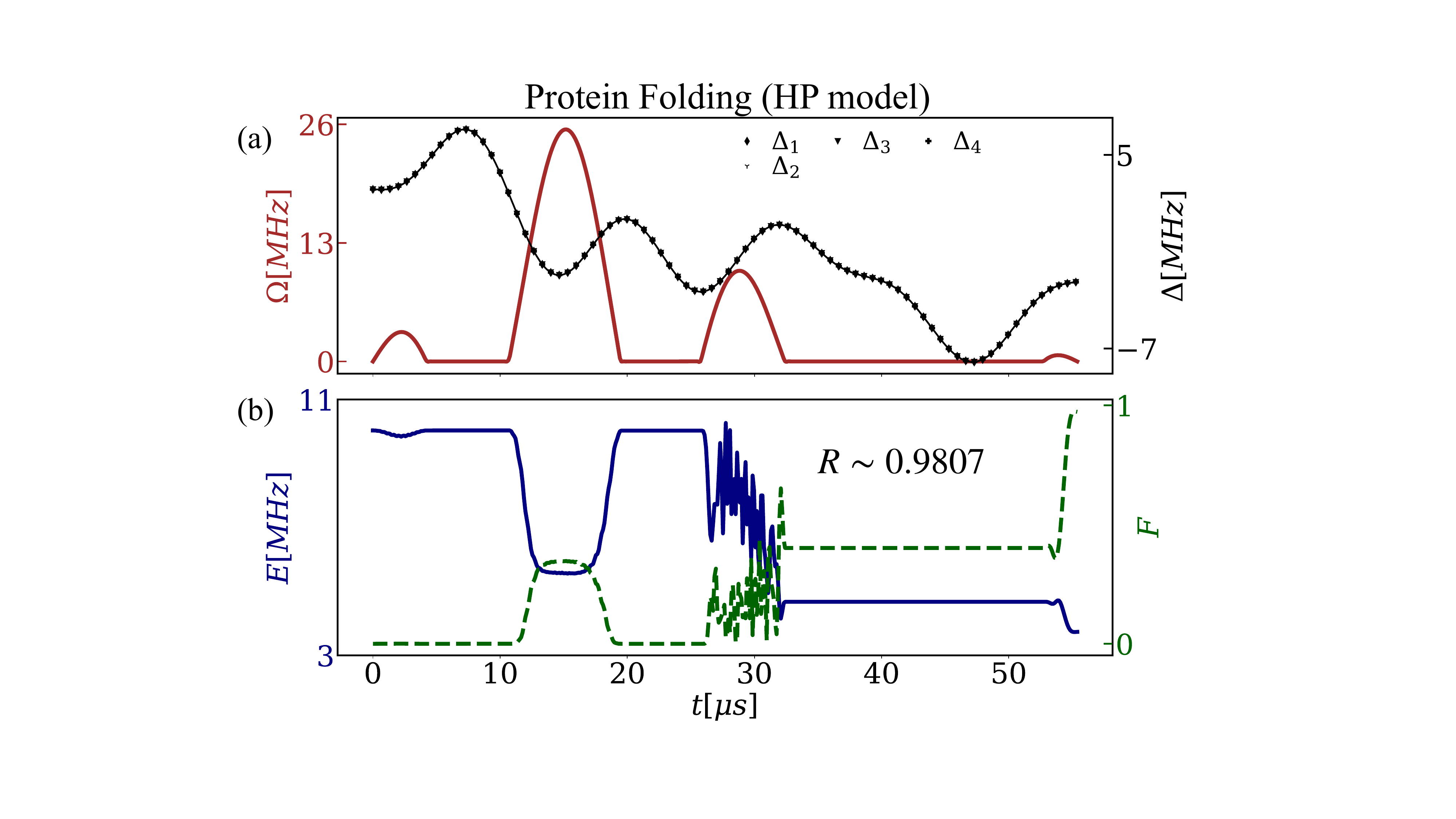}}
\caption{The optimal protocol for finding the solution of the toy model of the protein folding problem (Eq.~\ref{pfex}). [(a)] the optimal protocols for the Rabi frequency (solid dark red) and detunings (dotted black symbols) with time $t$. [(b)] the expectation value $E$ (solid blue curve), the fidelity $F$ (dashed green curve) and the approximation ratio $R$ (text). The solution is not exactly optimal in this case due to the problem having a relatively difficult optimization landscape to navigate.}
\label{f5}
\end{figure}

Finally, in Figs.~\ref{f6}~and~\ref{f5}, problems from vastly different fields are mapped to the Rydberg system and solved using our scheme to demonstrate the applicability of the protocol. Fig.~\ref{f6} shows the optimal solution for a binary clustering problem, which is relevant in machine learning, whereas a toy model of the protein folding problem is solved in Fig.~\ref{f5}. 
For the binary clustering problem, we consider a weighted graph on five nodes with an adjacency matrix
\begin{align}
    W =
\begin{pmatrix}
0 & 3 & 0 & 0 & 1 \\
3 & 0 & 2 & 0 & 0 \\
0 & 2 & 0 & 4 & 1 \\
0 & 0 & 4 & 0 & 2 \\
1 & 0 & 1 & 2 & 0
\end{pmatrix}.
\label{bcex}
\end{align}
The goal is to partition the nodes into two clusters. A standard formulation of this problem is the weighted max-Cut objective $\max_{x \in \{0,1\}^5} \sum_{i<j} W_{ij}(x_i \oplus x_j)$, which can be written in quadratic form as $\max \sum_{i<j} W_{ij}(x_i + x_j - 2x_i x_j)$. Since node \(5\) is weakly connected to the other nodes, it leads to multiple near-optimal partitions rather than a single dominant clustering. There are four distinct detuning values over five nodes ($\Delta_1 = \Delta_5$) as shown in Fig.~\ref{f6}(a), and the problem is solved exactly, which is depicted by the approximation ratio $R\sim 1$. This is because the Max-cut problem (binary clustering) maps natively to the Rydberg Hamiltonian with localized detuning, making it a relatively easier problem to solve due to the structure, even though it is still NP-hard.

On the other hand, a non-native problem, that is, protein folding, reach only sub-par fidelity (less than one) using the Rydberg annealer. Consider a short hydrophobic-polar (HP) sequence $\text{HHPH}, \quad H = (1,1,0,1),$ embedded on a lattice. Let \(x_{i,p} \in \{0,1\}\) indicate whether amino acid \(i\) occupies lattice position \(p\). The constraints to enforce a valid configuration are,
\begin{align}
   \sum_p x_{i,p} = 1 \quad \forall i, \qquad
\sum_i x_{i,p} \le 1 \quad \forall p. 
\label{pfex}
\end{align}
The objective rewarding non-consecutive hydrophobic contacts is $E = \sum_{\substack{i<j \\ |i-j|>1}} H_i H_j \, C_{ij} + \text{penalties}$,
where \(C_{ij} = 1\) if residues \(i\) and \(j\) are adjacent on the lattice.
This problem instance is already non-trivial; only three hydrophobic
residues are present, limiting the number of favorable contacts. The optimal
folding will attempt to bring the first and last residues together, but geometric
self-avoidance constraints restrict feasible configurations. Consequently,
the instance has several configurations with similar energies and a relatively
flat energy landscape. Thus, finding the optimal solution is very difficult, as also seen in Fig.~\ref{f5}(b) with $R \sim 0.98$ and fidelities only reaching close to $0.97$.

\begin{table}[t]
\tbl{Hardness parameter of the QUBO problems. The tolerance for near-degeneracy $\epsilon = 10^{-10}$. \label{ta1}}
{\tabcolsep13pt\begin{tabular}{@{}ccccccc@{}}
\colrule
Problems &$E_0$&$G_\Delta$ &$D_{opt}$ &$D_{E_1}$  &No. of & $\mathcal{HP}_{MI}$ \\
&&$= \bar{E}_1 - \bar{E}_0$&&&$\alpha: D_\alpha \geq D_{opt}/2$&\\
\colrule
Two-SAT & $-0.15$ & $0.30$ & $4$ & $4$ & $1$ & $27.25$ \\
\colrule
XOR-SAT & $0.30$ & $0.60$ & $6$ & $2$ & $1$& $1.13$ \\
\colrule
Two-XOR-SAT & $-0.15$ & $0.30$ & $2$ & $4$ & $2$& $64.52$ \\
\colrule
Set packing & $0.0$ & $0.60$ & $2$ & $4$ & $5$& $4.79$ \\
\colrule
QAP & $0.20$ & $0.24$ & $4$ & $8$ & $3$& $79.93$ \\
\colrule
Binary clustering & $-0.78$ & $0.24$ & $1$ & $2$ & $18$& $89.15$ \\
\colrule
Protein folding & $0.28$ & $0.08$ & $6$ & $4$ & $2$& $307.81$ 
\\
\botrule
\end{tabular}}
\end{table}

\subsection{Hardness of the problems}
The above examples collectively highlight several phenomena in discrete
optimization. The role of \emph{problem structure} as bipartite constraints (set packing) and graphs without cycles
(two-SAT) lead to tractable problems, whereas odd cycles or conflicting constraints create frustration. Many instances show \emph{degeneracy or many energetically low-lying excited subspaces} in the optimization landscape, contributing to the difficulty in solving them. For example, the quadratic assignment instance has multiple global optima, while the clustering and protein folding problems contain several near-optimal solutions due to the interaction structure. The other relevant factor is the \emph{energy landscapes}, some of which contain accessible optima, while others are flat or rugged, making optimization more challenging. Overall, the collection of problems demonstrates how small changes in constraint type or structure can alter the computational and qualitative nature of a problem.

We now use the hardness parameter, as defined in Sec.~\ref{HP}, that specifically takes these factors into account for the given prototypical problems, shown in Tab.~\ref{ta1}. The table shows that even if a general problem is NP-hard (set packing), there are instances that can be solved efficiently due to the structure of the problem. The hardness parameter characterizing the specific instances of XOR-SAT and set packing is easier as they have the largest gap ($G_\Delta = 0.60$) out of all the problems, whereas this gap plays a big role in making protein folding the most difficult one. This is also evident by the fidelity and the approximation ratio of the solution reached by the protocol. The hardness of the binary clustering stems from the optimization landscape being staggered, as it is indicated by $18$ threatening subspaces where the optimizer can get stuck, and the large number of first excited state configurations $D_{E_1} = 8$ in QAP contributes to the flatness of the landscape, which also is an issue for optimization. This parameter can thus be used for any QUBO problem as a tool to analyze any optimization method for small instances of the problem, as it requires the spectral information of the corresponding Ising Hamiltonian.

\section{Conclusions and Outlook}
\label{Conclusions}

In this work, we have shown that a variety of combinatorial optimization problems, including two-SAT, XOR-SAT, mixed-two-XOR-SAT, set packing, quadratic assignment, binary clustering, and protein folding, can be treated within a unified approach based on their QUBO representations. This is done by leveraging a direct mapping from QUBO to Ising spin models and implementing it on a Rydberg platform with local light shifts, as it was originally introduced in Ref.~\refcite{kapil}. The encoding enables a treatment of both polynomial-time solvable and NP-hard problems within the same physical framework in a resource-efficient manner.
Unlike blockade-based schemes that incur quadratic overhead in the number of atoms with respect to the size of the QUBO, our scheme has a linear overhead, making it suitable for near-term quantum devices. Furthermore, local control of the atoms through detunings provides a flexible method for implementing problem-specific constraints without requiring different encoding schemes.

The optimized annealing protocol, based on tunable detuning and Rabi frequency profiles, finds the ground state of the target Hamiltonian corresponding to all the problems. We use a hybrid method for optimization using a gradient-based optimizer that takes the system to lower energy levels, a gradient-free optimizer to take the system out of local minima, and finally a gradient-based optimizer to fine-tune the solution. Since it's a one-to-one encoding scheme, all the degenerate ground states are the solution to the problem. Further investigations are required to study the robustness of the protocols against noise and their performance for more complex instances \cite{saffman_quantum_2010,wu2022erasure,graham2019rydberg}. Based on the problem structure, a hardness parameter is defined to quantify the difficulty of solving a certain QUBO problem, which is then used to analyze the prototypical problems in this work. 
Overall, the work highlights the potential of Rydberg atom platforms as analog quantum optimizers capable of handling a variety of QUBO problems within a common framework, and can act as a benchmark for other gate-based quantum methods \cite{miessen2024benchmarking}. Future directions include improving robustness through noise-adaptive optimization techniques and quantifying the quantum interference as a resource for any advantage.

\section*{Acknowledgments}
This work is funded by the German Federal Ministry of Education and Research within the funding program “Quantum Technologies - from basic research to market” under Contract No. 13N16138.

\section*{ORCID}

\noindent Kapil Goswami - \url{https://orcid.org/0009-0007-7282-7970}

\noindent Peter Schmelcher - \url{https://orcid.org/0000-0002-2637-0937}

\bibliographystyle{ws-ijmpd}
\bibliography{bib_file.bib}

@article{Gal,
year = {1988},
month = {feb},
volume = {51},
number = {2},
pages = {143},
author = {T F Gallagher},
title = {Rydberg atoms},
journal = {Reports on Progress in Physics},
}

@inproceedings{nusslein2022algorithmic,
  title={Algorithmic QUBO formulations for k-SAT and hamiltonian cycles},
  author={N{\"u}{\ss}lein, Jonas and Gabor, Thomas and Linnhoff-Popien, Claudia and Feld, Sebastian},
  booktitle={Proceedings of the Genetic and Evolutionary Computation Conference Companion},
  pages={2240},
  year={2022}
}

@inproceedings{ayodele2022penalty,
  title={Penalty weights in QUBO formulations: Permutation problems},
  author={Ayodele, Mayowa},
  booktitle={European Conference on Evolutionary Computation in Combinatorial Optimization (Part of EvoStar)},
  pages={159},
  year={2022},
  organization={Springer}
}

@article{irback2022folding,
  title={Folding lattice proteins with quantum annealing},
  author={Irb{\"a}ck, Anders and Knuthson, Lucas and Mohanty, Sandipan and Peterson, Carsten},
  journal={Physical Review Research},
  volume={4},
  number={4},
  pages={043013},
  year={2022},
  publisher={APS}
}

@article{fraenkel1993complexity,
  title={Complexity of protein folding},
  author={Fraenkel, Aviezri S},
  journal={Bulletin of Mathematical Biology},
  volume={55},
  number={6},
  pages={1199},
  year={1993},
  publisher={Elsevier}
}

@inproceedings{schaefer1978complexity,
  title={The complexity of satisfiability problems},
  author={Schaefer, Thomas J},
  booktitle={Proceedings of the tenth annual ACM symposium on Theory of computing},
  pages={216},
  year={1978}
}

@article{compiani2013computational,
  title={Computational and theoretical methods for protein folding},
  author={Compiani, Mario and Capriotti, Emidio},
  journal={Biochemistry},
  volume={52},
  number={48},
  pages={8601},
  year={2013},
  publisher={ACS Publications}
}

@article{zhang2018binary,
  title={Binary multi-view clustering},
  author={Zhang, Zheng and Liu, Li and Shen, Fumin and Shen, Heng Tao and Shao, Ling},
  journal={IEEE Transactions on Pattern Analysis and Machine Intelligence},
  volume={41},
  number={7},
  pages={1774},
  year={2018},
  publisher={IEEE}
}

@book{cela2013quadratic,
  title={The Quadratic Assignment Problem: Theory and Algorithms},
  author={Cela, Eranda},
  volume={1},
  year={2013},
  publisher={Springer Science \& Business Media}
}

@article{browaeys2020many,
  title={Many-body physics with individually controlled Rydberg atoms},
  author={Browaeys, Antoine and Lahaye, Thierry},
  journal={Nature Physics},
  volume={16},
  number={2},
  pages={132},
  year={2020}
}

@article{kaufman2021quantum,
  title={Quantum science with optical tweezer arrays of ultracold atoms and molecules},
  author={Kaufman, Adam M and Ni, Kang-Kuen},
  journal={Nature Physics},
  volume={17},
  number={12},
  pages={1324},
  year={2021}
}

@article{zeiher2016many,
  title={Many-body interferometry of a Rydberg-dressed spin lattice},
  author={Zeiher, Johannes and Van Bijnen, Rick and Schau{\ss}, Peter and Hild, Sebastian and Choi, Jae-yoon and Pohl, Thomas and Bloch, Immanuel and Gross, Christian},
  journal={Nature Physics},
  volume={12},
  number={12},
  pages={1095},
  year={2016}
}

@article{gaetan2009observation,
  title={Observation of collective excitation of two individual atoms in the Rydberg blockade regime},
  author={Ga{\"e}tan, Alpha and Miroshnychenko, Yevhen and Wilk, Tatjana and Chotia, Amodsen and Viteau, Matthieu and Comparat, Daniel and Pillet, Pierre and Browaeys, Antoine and Grangier, Philippe},
  journal={Nature Physics},
  volume={5},
  number={2},
  pages={115},
  year={2009}
}

@article{low2012experimental,
  title={An experimental and theoretical guide to strongly interacting Rydberg gases},
  author={L{\"o}w, Robert and Weimer, Hendrik and Nipper, Johannes and Balewski, Jonathan B and Butscher, Bj{\"o}rn and B{\"u}chler, Hans Peter and Pfau, Tilman},
  journal={Journal of Physics B},
  volume={45},
  number={11},
  pages={113001},
  year={2012}
}

@article{saffman_quantum_2010,
  title={Quantum information with Rydberg atoms},
  author={Saffman, Mark and Walker, Thad G and M{\o}lmer, Klaus},
  journal={Reviews of Modern Physics},
  volume={82},
  number={3},
  pages={2313},
  year={2010},
  publisher={APS}
}

@article{hadfield_quantum_2019,
  title={From the quantum approximate optimization algorithm to a quantum alternating operator ansatz},
  author={Hadfield, Stuart and Wang, Zhihui and O’gorman, Bryan and Rieffel, Eleanor G and Venturelli, Davide and Biswas, Rupak},
  journal={Algorithms},
  volume={12},
  number={2},
  pages={34},
  year={2019},
  publisher={MDPI}
}

@article{farhi_quantum_2014,
  title={A quantum approximate optimization algorithm},
  author={Farhi, Edward and Goldstone, Jeffrey and Gutmann, Sam},
  journal={arXiv:1411.4028},
  year={2014}
}

@article{lucas_ising_2014,
  title={Ising formulations of many NP problems},
  author={Lucas, Andrew},
  journal={Frontiers in Physics},
  volume={2},
  pages={5},
  year={2014},
  publisher={Frontiers}
}

@article{evered2023high,
  title={High-fidelity parallel entangling gates on a neutral-atom quantum computer},
  author={Evered, Simon J and Bluvstein, Dolev and Kalinowski, Marcin and Ebadi, Sepehr and Manovitz, Tom and Zhou, Hengyun and Li, Sophie H and Geim, Alexandra A and Wang, Tout T and Maskara, Nishad and others},
  journal={Nature},
  volume={622},
  number={7982},
  pages={268},
  year={2023},
  publisher={Nature Publishing Group UK London}
}

@article{bluvstein2024logical,
  title={Logical quantum processor based on reconfigurable atom arrays},
  author={Bluvstein, Dolev and Evered, Simon J and Geim, Alexandra A and Li, Sophie H and Zhou, Hengyun and Manovitz, Tom and Ebadi, Sepehr and Cain, Madelyn and Kalinowski, Marcin and Hangleiter, Dominik and others},
  journal={Nature},
  volume={626},
  number={7997},
  pages={58},
  year={2024},
  publisher={Nature Publishing Group UK London}
}

@article{lee2026fundamental,
  title={Fundamental Challenges, Physical Implementations, and Integration Strategies for Ising Machines in Large-Scale Optimization Tasks},
  author={Lee, Hyunjun and Kim, Joon Pyo and Kim, Sanghyeon},
  journal={Advanced Electronic Materials},
  pages={e00682},
  year={2026},
  publisher={Wiley Online Library}
}

@article{mbeng2024quantum,
  title={The quantum Ising chain for beginners},
  author={Mbeng, Glen Bigan and Russomanno, Angelo and Santoro, Giuseppe E},
  journal={SciPost Physics Lecture Notes},
  pages={082},
  year={2024}
}

@article{kim_rydberg_2021,
  title={Rydberg quantum wires for maximum independent set problems},
  author={Kim, Minhyuk and Kim, Kangheun and Hwang, Jaeyong and Moon, Eun-Gook and Ahn, Jaewook},
  journal={Nature Physics},
  volume={18},
  number={7},
  pages={755},
  year={2022},
  publisher={Nature Publishing Group UK London}
}

@article{nguyen_quantum_2022,
  title={Quantum optimization with arbitrary connectivity using rydberg atom arrays},
  author={Nguyen, Minh-Thi and Liu, Jin-Guo and Wurtz, Jonathan and Lukin, Mikhail D and Wang, Sheng-Tao and Pichler, Hannes},
  journal={Physical Review X Quantum},
  volume={4},
  number={1},
  pages={010316},
  year={2023},
  publisher={APS}
}

@article{glover_tutorial_2019,
  title={A tutorial on formulating and using QUBO models},
  author={Glover, Fred and Kochenberger, Gary and Du, Yu},
  journal={arXiv:1811.11538},
  year={2018}
}

@article{barahona1988application,
  title={An application of combinatorial optimization to statistical physics and circuit layout design},
  author={Barahona, Francisco and Gr{\"o}tschel, Martin and J{\"u}nger, Michael and Reinelt, Gerhard},
  journal={Operations Research},
  volume={36},
  number={3},
  pages={493},
  year={1988},
  publisher={INFORMS}
}

@Inbook{karp1972complexity,
author={Karp, Richard M.},
editor={Miller, Raymond E.
and Thatcher, James W.
and Bohlinger, Jean D.},
title={Reducibility among Combinatorial Problems},
bookTitle={Complexity of Computer Computations: Proceedings of a Symposium on the Complexity of Computer Computations},
year={1972},
publisher={Springer US},
address={Boston, MA},
pages={85}
}

@article{bittel2022optimizing,
  title={Optimizing the depth of variational quantum algorithms is strongly QCMA-hard to approximate},
  author={Bittel, Lennart and Gharibian, Sevag and Kliesch, Martin},
  journal={arXiv:2211.12519},
  year={2022}
}

@article{bittel2021training,
  title={Training variational quantum algorithms is np-hard},
  author={Bittel, Lennart and Kliesch, Martin},
  journal={Physical Review Letters},
  volume={127},
  number={12},
  pages={120502},
  year={2021},
  publisher={APS}
}

@phdthesis{butenko2003maximum,
  title={Maximum Independent Set and Related Problems, with Applications},
  author={Butenko, Sergiy},
  year={2003},
  publisher={University of Florida}
}

@article{kadowaki1998quantum,
  title={Quantum annealing in the transverse Ising model},
  author={Kadowaki, Tadashi and Nishimori, Hidetoshi},
  journal={Physical Review E},
  volume={58},
  number={5},
  pages={5355},
  year={1998},
  publisher={APS}
}

@article{miessen2024benchmarking,
  title={Benchmarking digital quantum simulations above hundreds of qubits using quantum critical dynamics},
  author={Miessen, Alexander and Egger, Daniel J and Tavernelli, Ivano and Mazzola, Guglielmo},
  journal={Physical Review X Quantum},
  volume={5},
  number={4},
  pages={040320},
  year={2024},
  publisher={APS}
}

@inbook{karp1975computational,
  title={On the Computational Complexity of Combinatorial Problems},
  author={Karp, Richard M},
  journal={Networks},
  volume={5},
  number={1},
  pages={45},
  year={1975},
  publisher={Wiley Online Library}
}

@inproceedings{papadimitriou1988optimization,
  title={Optimization, approximation, and complexity classes},
  author={Papadimitriou, Christos and Yannakakis, Mihalis},
  booktitle={Proceedings of the Twentieth Annual ACM Symposium on Theory of Computing},
  pages={229},
  year={1988}
}

@article{hochba1997approximation,
  title={Approximation algorithms for NP-hard problems},
  author={Hochba, Dorit S},
  journal={ACM SIGACT News},
  volume={28},
  number={2},
  pages={40},
  year={1997},
  publisher={ACM New York, NY, USA}
}

@article{im2025accelerating,
  title={Accelerating Hybrid XOR $-$ CNF SAT Problems Natively with In-Memory Computing},
  author={Im, Haesol and B{\"o}hm, Fabian and Pedretti, Giacomo and Kushida, Noriyuki and Noori, Moslem and Valiante, Elisabetta and Zhang, Xiangyi and Yang, Chan-Woo and Bhattacharya, Tinish and Sheng, Xia and others},
  journal={arXiv:2504.06476},
  year={2025}
}

@incollection{vemuganti1998applications,
  title={Applications of set covering, set packing and set partitioning models: A survey},
  author={Vemuganti, Rao R},
  booktitle={Handbook of Combinatorial Optimization: Volume 1--3},
  pages={573},
  year={1998},
  publisher={Springer}
}

@article{dill2008protein,
  title={The protein folding problem},
  author={Dill, Ken A and Ozkan, S Banu and Shell, M Scott and Weikl, Thomas R},
  journal={Annual Review of Biophysics},
  volume={37},
  number={1},
  pages={289},
  year={2008},
  publisher={Annual Reviews}
}

@article{stillinger1993toy,
  title={Toy model for protein folding},
  author={Stillinger, Frank H and Head-Gordon, Teresa and Hirshfeld, Catherine L},
  journal={Physical Review E},
  volume={48},
  number={2},
  pages={1469},
  year={1993},
  publisher={APS}
}

@article{rodriguez2020clustering,
  title={Clustering improves the Goemans--Williamson approximation for the max-cut problem},
  author={Rodriguez-Fernandez, Angel E and Gonzalez-Torres, Bernardo and Menchaca-Mendez, Ricardo and Stadler, Peter F},
  journal={Computation},
  volume={8},
  number={3},
  pages={75},
  year={2020},
  publisher={MDPI}
}

@article{wang2023graph,
  title={Graph-collaborated auto-encoder hashing for multiview binary clustering},
  author={Wang, Huibing and Yao, Mingze and Jiang, Guangqi and Mi, Zetian and Fu, Xianping},
  journal={IEEE Transactions on Neural Networks and Learning Systems},
  volume={35},
  number={7},
  pages={10121},
  year={2023},
  publisher={IEEE}
}

@incollection{pothen1997graph,
  title={Graph partitioning algorithms with applications to scientific computing},
  author={Pothen, Alex},
  booktitle={Parallel Numerical Algorithms},
  pages={323},
  year={1997},
  publisher={Springer}
}

@inproceedings{codognet2022quantum,
  title={Quantum and digital annealing for the quadratic assignment problem},
  author={Codognet, Philippe and Diaz, Daniel and Abreu, Salvador},
  booktitle={2022 IEEE International Conference on Quantum Software (QSW)},
  pages={1},
  year={2022},
  organization={IEEE}
}

@article{commander2005survey,
  title={A survey of the quadratic assignment problem, with applications},
  author={Commander, Clayton W},
  year={2005}
}

@article{hazan2006complexity,
  title={On the complexity of approximating k-set packing},
  author={Hazan, Elad and Safra, Shmuel and Schwartz, Oded},
  journal={Computational Complexity},
  volume={15},
  number={1},
  pages={20},
  year={2006},
  publisher={Springer}
}

@article{ansotegui2025sat,
  title={SAT, gadgets, Max2XOR, and quantum annealers},
  author={Ans{\'o}tegui, Carlos and Levy, Jordi},
  journal={Quantum Information Processing},
  volume={24},
  number={10},
  pages={1},
  year={2025},
  publisher={Springer}
}

@article{king2019quantum,
  title={Quantum annealing amid local ruggedness and global frustration},
  author={King, James and Yarkoni, Sheir and Raymond, Jack and Ozfidan, Isil and King, Andrew D and Nevisi, Mayssam Mohammadi and Hilton, Jeremy P and McGeoch, Catherine C},
  journal={Journal of the Physical Society of Japan},
  volume={88},
  number={6},
  pages={061007},
  year={2019},
  publisher={The Physical Society of Japan}
}

@inproceedings{dietzfelbinger2010tight,
  title={Tight thresholds for cuckoo hashing via XORSAT},
  author={Dietzfelbinger, Martin and Goerdt, Andreas and Mitzenmacher, Michael and Montanari, Andrea and Pagh, Rasmus and Rink, Michael},
  booktitle={International Colloquium on Automata, Languages, and Programming},
  pages={213},
  year={2010},
  organization={Springer}
}

@article{jordan2025optimization,
  title={Optimization by decoded quantum interferometry},
  author={Jordan, Stephen P and Shutty, Noah and Wootters, Mary and Zalcman, Adam and Schmidhuber, Alexander and King, Robbie and Isakov, Sergei V and Khattar, Tanuj and Babbush, Ryan},
  journal={Nature},
  volume={646},
  number={8086},
  pages={831},
  year={2025},
  publisher={Nature Publishing Group UK London}
}

@book{ganai2007sat,
  title={SAT-based Scalable Formal Verification Solutions},
  author={Ganai, Malay K and Gupta, Aarti},
  year={2007},
  publisher={Springer}
}

@article{cho1998fast,
  title={Fast approximation algorithms on maxcut, k-coloring, and k-color ordering for VLSI applications},
  author={Cho, Jun-Dong and Raje, Salil and Sarrafzadeh, Majid},
  journal={IEEE Transactions on Computers},
  volume={47},
  number={11},
  pages={1253},
  year={1998},
  publisher={IEEE}
}

@article{peruzzo2014variational,
  title={A variational eigenvalue solver on a photonic quantum processor},
  author={Peruzzo, Alberto and McClean, Jarrod and Shadbolt, Peter and Yung, Man-Hong and Zhou, Xiao-Qi and Love, Peter J and Aspuru-Guzik, Al{\'a}n and O’brien, Jeremy L},
  journal={Nature Communications},
  volume={5},
  number={1},
  pages={4213},
  year={2014},
  publisher={Nature Publishing Group UK London}
}

@article{hochbaum1998analysis,
  title={Analysis of the greedy approach in problems of maximum k-coverage},
  author={Hochbaum, Dorit S and Pathria, Anu},
  journal={Naval Research Logistics},
  volume={45},
  number={6},
  pages={615},
  year={1998},
  publisher={Wiley Online Library}
}

@article{perelshtein2022practical,
  title={Practical application-specific advantage through hybrid quantum computing},
  author={Perelshtein, Michael and Sagingalieva, Asel and Pinto, Karan and Shete, Vishal and Pakhomchik, Alexey and Melnikov, Artem and Neukart, Florian and Gesek, Georg and Melnikov, Alexey and Vinokur, Valerii},
  journal={arXiv:2205.04858},
  year={2022}
}

@inproceedings{scheideler2008log,
  title={An O (log n) dominating set protocol for wireless ad-hoc networks under the physical interference model},
  author={Scheideler, Christian and Richa, Andrea and Santi, Paolo},
  booktitle={Proceedings of the 9th ACM International Symposium on Mobile ad hoc Networking and Computing},
  pages={91},
  year={2008}
}

@article{devoret2004superconducting,
  title={Superconducting qubits: A short review},
  author={Devoret, Michel H and Wallraff, Andreas and Martinis, John M},
  journal={arXiv:cond-mat/0411174v1},
  year={2004}
}

@article{bruzewicz2019trapped,
  title={Trapped-ion quantum computing: Progress and challenges},
  author={Bruzewicz, Colin D and Chiaverini, John and McConnell, Robert and Sage, Jeremy M},
  journal={Applied Physics Reviews},
  volume={6},
  number={2},
  pages={021314},
  year={2019},
  publisher={AIP Publishing LLC}
}

@article{harter2014long,
  title={Long-term drifts of stray electric fields in a Paul trap},
  author={H{\"a}rter, Arne and Kr{\"u}kow, Artjom and Brunner, Andreas and Hecker Denschlag, Johannes},
  journal={Applied Physics B},
  volume={114},
  pages={275},
  year={2014},
  publisher={Springer}
}

@article{gross2017quantum,
  title={Quantum simulations with ultracold atoms in optical lattices},
  author={Gross, Christian and Bloch, Immanuel},
  journal={Science},
  volume={357},
  number={6355},
  pages={995},
  year={2017},
  publisher={American Association for the Advancement of Science}
}

@article{urban2009observation,
  title={Observation of Rydberg blockade between two atoms},
  author={Urban, E and Johnson, Todd A and Henage, T and Isenhower, L and Yavuz, DD and Walker, TG and Saffman, M},
  journal={Nature Physics},
  volume={5},
  number={2},
  pages={110},
  year={2009},
  publisher={Nature Publishing Group UK London}
}

@article{feng2009lifetime,
  title={Lifetime measurement of ultracold caesium Rydberg states},
  author={Feng, Zhi-Gang and Zhang, Lin-Jie and Zhao, Jian-Ming and Li, Chang-Yong and Jia, Suo-Tang},
  journal={Journal of Physics B},
  volume={42},
  number={14},
  pages={145303},
  year={2009},
  publisher={IOP Publishing}
}

@book{wolsey1999integer,
  title={Integer and Combinatorial Optimization},
  author={Wolsey, Laurence A and Nemhauser, George L},
  year={1999},
  publisher={John Wiley \& Sons},
  doi={10.1002/9781118627372}
}

@article{gu1997algorithms,
  title={Algorithms for the satisfiability (SAT) problem: A survey},
  author={Gu, Jun and Purdom, Paul W and Franco, John and Wah, Benjamin W},
  journal={DIMACS Series in Discrete Mathematics and Theoretical Computer Science},
  volume={35},
  pages={19},
  year={1997}
}

@article{pirnay2024principle,
  title={An in-principle super-polynomial quantum advantage for approximating combinatorial optimization problems via computational learning theory},
  author={Pirnay, Niklas and Ulitzsch, Vincent and Wilde, Frederik and Eisert, Jens and Seifert, Jean-Pierre},
  journal={Science Advances},
  volume={10},
  number={11},
  pages={eadj5170},
  year={2024},
  publisher={American Association for the Advancement of Science}
}

@article{goswami2025qudit,
  title={Qudit-based scalable quantum algorithm for solving the integer programming problem},
  author={Goswami, Kapil and Schmelcher, Peter and Mukherjee, Rick},
  journal={arXiv:2508.13906},
  year={2025}
}

@article{goswami2026solving,
  title={Solving the travelling salesman problem using Bloch sphere encoding},
  author={Goswami, Kapil and Anekonda Veereshi, Gagan and Schmelcher, Peter and Mukherjee, Rick},
  journal={Quantum Science and Technology},
  volume={11},
  number={1},
  pages={015007},
  year={2026},
  publisher={IOP Publishing}
}

@article{mandal2023review,
  title={A review of classical methods and Nature-Inspired Algorithms (NIAs) for optimization problems},
  author={Mandal, Pawan Kumar},
  journal={Results in Control and Optimization},
  volume={13},
  pages={100315},
  year={2023},
  publisher={Elsevier},
  doi={10.1016/j.rico.2023.100315}
}

@article{abbas2024challenges,
  title={Challenges and opportunities in quantum optimization},
  author={Abbas, Amira and Ambainis, Andris and Augustino, Brandon and B{\"a}rtschi, Andreas and Buhrman, Harry and Coffrin, Carleton and Cortiana, Giorgio and Dunjko, Vedran and Egger, Daniel J and Elmegreen, Bruce G and others},
  journal={Nature Reviews Physics},
  volume = {6},
  pages={718},
  year={2024},
  publisher={Nature Publishing Group UK London},
  doi={10.1038/s42254-024-00770-9}
}

@article{vsibalic2017arc,
  title={ARC: An open-source library for calculating properties of alkali Rydberg atoms},
  author={{\v{S}}ibali{\'c}, Nikola and Pritchard, Jonathan D and Adams, Charles S and Weatherill, Kevin J},
  journal={Computer Physics Communications},
  volume={220},
  pages={319},
  year={2017},
  publisher={Elsevier}
}

@article{ebadi2022quantum,
  title={Quantum optimization of maximum independent set using Rydberg atom arrays},
  author={Ebadi, Sepehr and Keesling, Alexander and Cain, Madelyn and Wang, Tout T and Levine, Harry and Bluvstein, Dolev and Semeghini, Giulia and Omran, Ahmed and Liu, J-G and Samajdar, Rhine and others},
  journal={Science},
  volume={376},
  number={6598},
  pages={1209},
  year={2022},
  publisher={American Association for the Advancement of Science}
}

@inbook{vazirani2001approximation,
  title={Approximation Algorithms},
  author={Vazirani, Vijay V},
  volume={1},
  year={2001},
  pages={2},
  publisher={Springer}
}

@article{wu2022erasure,
  title={Erasure conversion for fault-tolerant quantum computing in alkaline earth Rydberg atom arrays},
  author={Wu, Yue and Kolkowitz, Shimon and Puri, Shruti and Thompson, Jeff D},
  journal={Nature Communications},
  volume={13},
  number={1},
  pages={4657},
  year={2022},
  publisher={Nature Publishing Group UK London}
}

@article{graham2019rydberg,
  title={Rydberg-mediated entanglement in a two-dimensional neutral atom qubit array},
  author={Graham, TM and Kwon, M and Grinkemeyer, B and Marra, Z and Jiang, X and Lichtman, MT and Sun, Y and Ebert, M and Saffman, M},
  journal={Physical Review Letters},
  volume={123},
  number={23},
  pages={230501},
  year={2019},
  publisher={APS}
}

@article{hare2013survey,
  title={A survey of non-gradient optimization methods in structural engineering},
  author={Hare, Warren and Nutini, Julie and Tesfamariam, Solomon},
  journal={Advances in Engineering Software},
  volume={59},
  pages={19},
  year={2013},
  publisher={Elsevier}
}

@book{conn2009introduction,
  title={Introduction to Derivative-free Optimization},
  author={Conn, Andrew R and Scheinberg, Katya and Vicente, Luis N},
  year={2009},
  publisher={SIAM}
}

@book{polak2012optimization,
  title={Optimization: Algorithms and Consistent Approximations},
  author={Polak, Elijah},
  volume={124},
  year={2012},
  publisher={Springer Science \& Business Media}
}

@book{hasdorff1976gradient,
  title={Gradient Optimization and Nonlinear Control},
  author={Hasdorff, Lawrence},
  year={1976},
  publisher = {Wiley}
}

@article{fletcher1970new,
  title={A new approach to variable metric algorithms},
  author={Fletcher, Roger},
  journal={The Computer Journal},
  volume={13},
  number={3},
  pages={317},
  year={1970},
  publisher={Oxford University Press}
}

@article{broyden1970convergence,
  title={The convergence of a class of double-rank minimization algorithms 1. general considerations},
  author={Broyden, Charles George},
  journal={IMA Journal of Applied Mathematics},
  volume={6},
  number={1},
  pages={76},
  year={1970},
  publisher={Oxford University Press}
}

@article{goldfarb1970family,
  title={A family of variable metric updates derived by variational means, v. 24},
  author={Goldfarb, D},
  journal={Mathematics of Computation},
  year={1970}
}

@article{shanno1970conditioning,
  title={Conditioning of quasi-Newton methods for function minimization},
  author={Shanno, David F},
  journal={Mathematics of Computation},
  volume={24},
  number={111},
  pages={647},
  year={1970}
}

@article{nelder1965simplex,
  title={A simplex method for function minimization},
  author={Nelder, John A and Mead, Roger},
  journal={The Computer Journal},
  volume={7},
  number={4},
  pages={308},
  year={1965},
  publisher={Oxford University Press}
}

@article{rabitz,
	author = {Rabitz, Herschel and de Vivie-Riedle, Regina and Motzkus, Marcus and Kompa, Karl},
	title = {Whither the Future of Controlling Quantum Phenomena?},
	volume = {288},
	number = {5467},
	pages = {824},
	year = {2000},
	doi = {10.1126/science.288.5467.824},
	publisher = {American Association for the Advancement of Science},
	issn = {0036-8075},
	URL = {https://science.sciencemag.org/content/288/5467/824},
	journal = {Science}
}

@article{Mukherjee_2020,
	doi = {10.1088/1367-2630/ab8677},
	url = {https://doi.org/10.1088%2F1367-2630%2Fab8677},
	year = 2020,
	publisher = {{IOP} Publishing},
	volume = {22},
	number = {7},
	pages = {075001},
	author = {Rick Mukherjee and Fr{\'{e}}d{\'{e}}ric Sauvage and Harry Xie and Robert Löw and Florian Mintert},
	title = {Preparation of ordered states in ultra-cold gases using Bayesian optimization},
	journal = {New Journal of Physics},
}

@article{Mukherjee2,
	title = {Bayesian Optimal Control of Greenberger-Horne-Zeilinger States in Rydberg Lattices},
	author = {Mukherjee, Rick and Xie, Harry and Mintert, Florian},
	journal = {Physical Review Letters},
	volume = {125},
	issue = {20},
	pages = {203603},
	numpages = {6},
	year = {2020},
	publisher = {American Physical Society},
	doi = {10.1103/PhysRevLett.125.203603},
	url = {https://link.aps.org/doi/10.1103/PhysRevLett.125.203603}
}

@article{kelly2014optimal,
  title={Optimal quantum control using randomized benchmarking},
  author={Kelly, Julian and Barends, Rami and Campbell, Brooks and Chen, Yu and Chen, Zijun and Chiaro, Ben and Dunsworth, Andrew and Fowler, Austin G and Hoi, I-C and Jeffrey, Evan and others},
  journal={Physical Review Letters},
  volume={112},
  number={24},
  pages={240504},
  year={2014},
  publisher={APS}
}

@article{li2017hybrid,
	title = {Hybrid Quantum-Classical Approach to Quantum Optimal Control},
	author = {Li, Jun and Yang, Xiaodong and Peng, Xinhua and Sun, Chang-Pu},
	journal = {Physical Review Letters},
	volume = {118},
	issue = {15},
	pages = {150503},
	numpages = {5},
	year = {2017},
	publisher = {American Physical Society},
	doi = {10.1103/PhysRevLett.118.150503},
	url = {https://link.aps.org/doi/10.1103/PhysRevLett.118.150503}
}

@article{PhysRevA.91.052306,
	title = {Robust and efficient in situ quantum control},
	author = {Ferrie, Christopher and Moussa, Osama},
	journal = {Physical Review A},
	volume = {91},
	issue = {5},
	pages = {052306},
	numpages = {8},
	year = {2015},
	publisher = {American Physical Society},
	doi = {10.1103/PhysRevA.91.052306},
	url = {https://link.aps.org/doi/10.1103/PhysRevA.91.052306}
}

@article{CRAB1,
	title = {Chopped random-basis quantum optimization},
	author = {Caneva, Tommaso and Calarco, Tommaso and Montangero, Simone},
	journal = {Physical Review A},
	volume = {84},
	issue = {2},
	pages = {022326},
	numpages = {9},
	year = {2011},
	publisher = {American Physical Society},
	doi = {10.1103/PhysRevA.84.022326},
	url = {https://link.aps.org/doi/10.1103/PhysRevA.84.022326}
}

@article{maurer2011spatial,
  title={What spatial light modulators can do for optical microscopy},
  author={Maurer, Christian and Jesacher, Alexander and Bernet, Stefan and Ritsch-Marte, Monika},
  journal={Laser \& Photonics Rev.},
  volume={5},
  number={1},
  pages={81--101},
  year={2011},
  publisher={Wiley Online Library}
}

@article{kapil,
  title = {Solving optimization problems with local light-shift encoding on Rydberg quantum annealers},
  author = {Goswami, Kapil and Mukherjee, Rick and Ott, Herwig and Schmelcher, Peter},
  journal = {Physical Review Research},
  volume = {6},
  year = {2024},
  month = {Apr},
  pages={023031},
  publisher = {Amer. Phys. Soc.},
  doi={10.1103/PhysRevResearch.6.023031}
}

@article{goswami2024integer,
  title={Integer programming using a single atom},
  author={Goswami, Kapil and Schmelcher, Peter and Mukherjee, Rick},
  journal={Quantum Science and Technology},
  volume={9},
  number={4},
  pages={045016},
  year={2024},
  publisher={IOP Publishing},
  doi={10.1088/2058-9565/ad6735}
}

\end{document}